\documentclass[11pt]{article}

\usepackage{xcolor}
\definecolor{pblue}{HTML}{0395DE}

\usepackage[colorlinks=true,
            linkcolor=pblue,
            urlcolor=pblue,
            citecolor=pblue,
            breaklinks=true]{hyperref}
\usepackage{breakurl} 
\usepackage[a4paper, left=3cm, right=3cm, top=3cm, bottom=3cm]{geometry}
\RequirePackage{amsmath,amsfonts,amssymb,amsthm}
\usepackage{mathtools}
\usepackage{xcolor}
\usepackage{mathrsfs} 
\usepackage{natbib}
\usepackage{setspace}
\usepackage{dsfont}
\usepackage{caption}
\usepackage{amsmath}

\usepackage{algorithm}
\usepackage{algorithmic}

\usepackage{caption}
\usepackage{chronosys}

\usepackage{mdframed}
\mdfdefinestyle{ndef-frame}{
linewidth=2pt, %
linecolor= pblue, %
backgroundcolor= gray!5,
topline=false, %
bottomline=false, %
rightline=false,%
leftmargin=0pt, %
innerleftmargin=0.2em, %
innerrightmargin=1em, 
rightmargin=0pt, %
innertopmargin=0em,%
innerbottommargin=1em, %
splittopskip=\topskip, %
}%
\surroundwithmdframed[style=ndef-frame]{ndef}
\theoremstyle{definition-style}

\mdfdefinestyle{lemma-frame}{
linewidth=1pt, %
linecolor= pblue, %
backgroundcolor= gray!5,
topline=true, %
bottomline=true, %
rightline=true,%
leftmargin=0pt, %
innerleftmargin=0.2em, %
innerrightmargin=1em, 
rightmargin=0pt, %
innertopmargin=0em,%
innerbottommargin=1em, %
splittopskip=\topskip, %
}%
\surroundwithmdframed[style=lemma-frame]{lemma}
\theoremstyle{lemma-style}

\hypersetup{pdfauthor = {Antoine Barbieri}}

\usepackage{authblk}

\newcommand{\ind}[1]{\mathds{1}_{#1}}

\newcommand{\R}{\mathds R}

\title{\vspace{-.5cm} Asymmetric Laplace distribution regression model for fitting heterogeneous longitudinal response \vspace{.5cm}}
\date{\vspace{-5ex}}
\author[1]{Antoine Barbieri\thanks{Corresponding author: \href{mailto:antoine.barbieri@u-bordeaux.fr}{\texttt{antoine.barbieri@u-bordeaux.fr}}}}
\author[2]{Angelo Alcaraz}
\author[1]{Mouna Abed}
\author[1,3]{Hugues de Courson}
\author[1]{H\'el\`ene Jacqmin-Gadda}
\affil[1]{Univ. Bordeaux, Bordeaux Population Health center, Inserm U1219, Bordeaux, France}
\affil[2]{Univ Bretagne Sud, CNRS UMR 6205, LMBA, France}
\affil[3]{Anesthesiology and Critical Care Department, Bordeaux University Hospital, Bordeaux, France}
\begin{document}

\maketitle

\begin{abstract}
The systematic collection of longitudinal data is very common in practice, making mixed models widely used. 
Most developments around these models focus on modeling the mean trajectory of repeated measurements, typically under the assumption of homoskedasticity. However, as data become increasingly rich through intensive collection over time, these models can become limiting and may introduce biases in analysis. In fact, such data are often heterogeneous, with the presence of outliers, heteroskedasticity, and asymmetry in the distribution of individual measurements.
Therefore, ignoring these characteristics can lead to biased modeling results.
In this work, we propose a mixed-effect distributional regression model based on the asymmetric Laplace distribution to: (1) address the presence of outliers, heteroskedasticity, and asymmetry in longitudinal measurements; (2) model the entire individual distribution of the heterogeneous longitudinal response over time, rather than just its conditional expectation; and (3) give a more comprehensive evaluation of the impact of covariates on the distribution of the responses through meaningful indicator.
A Bayesian estimation procedure is presented.
In order to choose between two distributional regression models, we also propose a new model selection criterion for longitudinal data. It measures the proximity between the individual distribution estimated by the model and the empirical individual distribution of the data over time, using a set of quantiles.
The estimation procedure and the selection criterion are validated in a simulation study and the proposed model is compared to a distributional regression mixed model based on the Gaussian distribution and a location-scale linear quantile mixed model.
Finally, the proposed model is applied to analyze blood pressure over time for hospitalized patients in the intensive care unit.
\end{abstract}


\section{Introduction}

Repeated measurements during patient follow-up enable individualized monitoring and management, and anticipation of possible events or deterioration in health status. In intensive care units, many biomarkers are measured very frequently to monitor the vital status of patients, to ensure that drug administration is effective, and to avoid undesirable events. Blood pressure is a key biological marker that is carefully regulated, often by medication, to maintain it within a specific range and ensure the proper functioning of vital organs. In this case, it is interesting to study the factors that influence the evolution of the biomarker, as well as the variability or asymmetry of its measurements, which are likely to affect patient vital status. Although all patients are closely monitored in the same way, the intensive collection of longitudinal data now provides a valuable source of information on biomarkers, both in terms of their evolution over time and their individual heterogeneity (central value, extreme values, heteroskedasticity, asymmetry...). \\

To assess the influence of factors on longitudinal responses, mixed models are particularly suitable and popular tools. The introduction of random effects makes it possible to take into account the correlation between observations from the same patient, to study the overall influence of factors (on the population level), and also to fit the response of interest at the individual level.
But classical linear mixed models (based on least-square optimization) focus on the mean behavior and study how covariates affect it, assuming that residual variability is the same for all measurements whatever the individual (\textit{i.e.} the homoscedasticity assumption).
However, longitudinal data often show variability that changes over time, is influenced by covariates, and can differ between individuals \citep{courcoul_SIM_2025}. Hence, assuming homoskedasticity may introduce bias in estimating the conditional distribution of the response variable and affect statistical inference \citep{Heller_gamlss_2022}. Furthermore, interest can also be in modeling residual marker variability \citep{courcoul_SIM_2025,alcaraz_lqmm_2025}, or other characteristics of the distribution other than simply mean \citep{Kneib_rage_2023}, as they may be a source of undesirable events.

In order to take into account the presence of heteroskedasticity in the data and to assess the impact of covariates on residual variability, generalized additive models for location, scale and shape (GAMLSS) have been proposed. 
GAMLSS models extend the classical regression framework by relaxing the assumption that the response variable follows a distribution from the exponential family \citep{2007_jss_gamlss}. They allow the use of a wide range of distributions, including continuous or discrete laws that exhibit strong skewness and/or marked flatness. Unlike traditional models that are limited to modeling the mean (or a location parameter), GAMLSS enable the simultaneous modeling of multiple distribution parameters (such as location, scale, or shape) through linear or non-linear, parametric or additive non-parametric functions, and can also incorporate explanatory variables and/or random effects. As such, GAMLSS are particularly well suited for analyzing data in which the response variable does not follow an exponential family distribution, for example, in the presence of skewed data or when the dispersion or shape parameters vary with covariates.
GAMLSS is a parametric distributional regression approach in which the response variable can have any parametric distribution and all its parameters can depend on explanatory variables and/or random effects \citep{Rigby_2021,Kneib_rage_2023}. 
For data analysis, this distributional approach offers a more realistic and richer framework than more traditional parameter regressions, and it also offers the possibility of obtaining a much more complete understanding of the measured behavior.

For fitting longitudinal data in the presence of heteroskedasticity, the location-scale mixed model (LSMM) based on the Gaussian assumption was developed \citep{hedeker_2008,courcoul_SIM_2025}. In their work, \citet{courcoul_SIM_2025} assumed that the response variable was normally distributed and adjusted both the location parameter and the scale parameter of the distribution for time, covariates, and subject-specific random effects. A major advantage of this model lies in its interpretation. In fact, the location parameter corresponds to the expectation, and the scale parameter corresponds to the residual variance. Thus, modeling these two parameters by explanatory variables and subject-specific random effects enables to study directly the impact of the explanatory variables on both the central behavior and the dispersion of the response distribution, at both population level (via fixed effects) and individual level (via random effects). Since the Gaussian distribution is entirely defined by its two parameters, LSMM falls within the scope of parametric distributional regression (or GAMLSS).
However, Gaussian assumption becomes inappropriate or restrictive if the data present an asymmetric distribution or outliers \citep{Stasinopoulos_2017}. 
Both situations can lead to biases in parameter estimation. 
Moreover, the other distributions available within the GAMLSS framework to handle skewness do not offer a direct and intuitive interpretation as the LSMM. For instance, the location parameter does not correspond exactly to a classical and intuitive statistic of central behavior such as the mean, median, or mode. \\

An alternative in the presence of asymmetry, outliers, and heteroskedasticity is quantile regression \citep{koenker_book_2005}. 
Quantile regression fits the conditional quantiles of a response variable based on a set of covariates, offering a more comprehensive view of the entire conditional distribution compared to classical linear regression, which focuses solely on the conditional mean. Indeed, quantile regression makes it possible to account for different variations in the relationship between the response variable and the explanatory variables across the conditional response distribution. This method is thus particularly useful in situations involving skewness, fat-tails, outliers, or heteroskedasticity \citep{petrella_2019}. When the interest is the central behavior of the distribution, modeling the median conditionally on the explanatory variables (and random effects for longitudinal data) rather than the expectation is more appropriate, as the median is known to be more robust than the mean to outliers.

For heterogeneous data, fitting different conditional quantiles of the response distribution enables the distribution to be explored. However, a regression model is estimated independently for each quantile of interest, which leads to two problems. The first is to use the same data several times to estimate different models to approximate the distribution of interest. The second is that independent estimates of the different quantiles may lead to theoretical violations, where the evolution of two distinct quantiles may intersect. For longitudinal data, this can lead to trajectories for different quantiles crossing each other, which is impossible in practice. Although multiple quantiles can be estimated separately, it therefore seems relevant to estimate them simultaneously \citep{liu_2009,cho_2017} or to consider a specific function that links the parameters of the different quantile regression \citep{frumento_2021}. In particular, there are two advantages of simultaneous estimation: to gain better accuracy of estimation using shared strength among them rather than individually estimated quantile functions, and to incorporate simultaneous non-crossing constraints of quantile regression functions in joint estimation \citep{liu_2011}. 
The disadvantage of quantile regression is that information about the distribution comes from the different modeled quantiles. This complicates statistical inference and interpretation, and studying the impact of covariates on the shape of the distribution of interest is neither easy nor direct. \\

In this work, we propose a distributional regression model for a heterogeneous longitudinal response at the crossroads of quantile regression and GAMLSS. It is a GAMLSS in which the response variable given individual random effects is assumed to follow an asymmetric Laplace ($\mathcal{AL}$) distribution. This distribution is naturally used in quantile regression when the estimation procedure is based on likelihood \citep{yu_bayesian_2001,geraci_2007,farcomeni_2015,yang_2019,petrella_2019}. 
It is defined by three parameters: a location parameter, a scale parameter, and a skewness parameter \citep{kotz_laplace_2012,yu_adl_2005}. The location parameter is the mode of the distribution that guaranties the evaluation of the central behavior of the distribution, and the scale and the skewness parameters are used to manage the dispersion and the asymmetry around the mode present in the data, respectively.
Consequently, this model combines the advantages of GAMLSS and quantile regression. Indeed, the adjustment of the distribution of the response variable directly takes into account asymmetry and heteroscedasticity by modeling the parameters as a function of the explanatory variables and random effects. In addition, it also inherits the robustness of the estimates to outliers highlighted in quantile regression.
Finally, since the $\mathcal{AL}$ distribution has an explicit distribution function, the quantiles of the distribution can be easily computed in a closed form.

The paper is organized as follows.
After introducing the asymmetric Laplace distribution and some of its properties, Section 2 details the proposed model and its particular cases, the proposed estimation procedure and a new model selection criterion. 
A simulation study to validate the estimation procedure and the selection criterion is described in Section 3. 
In Section 4, the approach is illustrated on blood pressure data collected in patients admitted to intensive care after subarachnoid hemorrhage at the Bordeaux University Hospital. 
The last section reviews the proposed work, its advantages and limitations.

\section{Asymmetric Laplace distribution regression model}

\subsection{Asymmetric Laplace distribution}

\begin{figure}[!t]
\begin{center}
\includegraphics[scale=0.85]{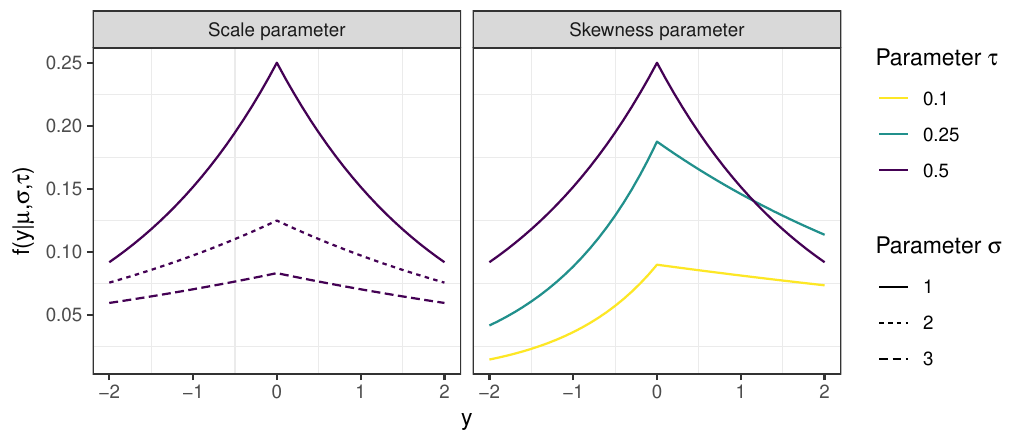}
\end{center}
\caption{Representation of $\mathcal{AL}$ density distribution $f\left(y\vert \mu=0,\sigma,\tau\right)$ according to different values of the skewness parameter $\tau$ (on right panel with $\sigma=1$) and the scale parameter $\sigma$ (on left panel with $\tau=0.5$). }\label{fig_ADL_rho}
\end{figure}

Assuming that $Y$ is a continuous random variable such as $Y\sim\mathcal{AL}\left(\mu,\sigma,\tau\right)$, its probability density function is
\begin{equation*}\label{eq:ALD_density}
f_Y\left(y \right) = \frac{\tau\left(1-\tau\right)}{\sigma} \exp\left\{ -\rho_\tau\left( \frac{y-\mu}{\sigma}\right)\right\},
\end{equation*}
where $\tau\in\left]0;1\right[$ is the skewness parameter, $\sigma\in\R^{+}_{*}$ is the scale parameter, and $\mu\in\R$ is the location parameter.  
The location parameter $\mu$ corresponds to both the mode and the quantile with order $\tau$ of the distribution. The scale and the skewness parameters manage dispersion and the asymmetry around the mode of the distribution.
Figure \ref{fig_ADL_rho} shows the impact of both skewness and scale parameters on the shape of the $\mathcal{AL}$ distribution. 
Note that $\rho_\tau\left( . \right)$ is the quantile loss function \citep{koenker_1978} defined for $v\in\R$ by:
\begin{eqnarray}\label{eq:rho}
\rho_\tau\left( v \right) & = & v \left( \tau-\ind{ v<0 } \right) \nonumber\\
& = & \left( \tau \ind{ v\geq 0 } + \left(1-\tau\right) \ind{ v< 0 } \right) |v| 
\end{eqnarray}
An interesting property of the $\mathcal{AL}$ distribution is that its cumulative distribution function is explicit \citep{yu_adl_2005}:
\begin{equation}\label{eq:ALD_cum_distr}
F_Y\left(y \right) = 
\left( \tau \exp\left(\frac{1-\tau}{\sigma}\left(y-\mu\right) \right) \right)\mathds{1}_{y\leq\mu} + 
\left( \tau-\left( 1-\tau \right) \exp\left(\frac{1-\tau}{\sigma}\left(y-\mu\right) \right) \right)\mathds{1}_{y > \mu}
\end{equation}
Moreover, the quantile of order $p$, denoted $Q_Y(p)$, of an $\mathcal{AL}$ distribution has a closed form defined as a function of the three parameters \citep{yu_adl_2005,kotz_laplace_2012}. 
Consider $p\in ]0,1[$, $Q_Y(p)$ is defined by $\Pr\left(Y\leq Q_Y(p) \right) = p$. Then
\begin{equation}\label{eq:ALD_quantile_fct}
Q_Y(p) = \left( \mu + \frac{\sigma}{1-\tau}\log\left(\frac{p}{\tau}\right) \right)\mathds{1}_{p\leq \tau} + \left( \mu-\frac{\sigma}{\tau} \log\left( \frac{1-p}{1-\tau} \right) \right)\mathds{1}_{p > \tau} .
\end{equation}
Finally, the variance and the expectation have also closed-form expressions given parameters:
\begin{equation}\label{eq:var_exp}
    \mathbb{V}ar(Y)=\sigma^2 \frac{\tau^2+(1-\tau)^2}{\tau^2(1-\tau)^2}
    \quad \text{ and } \quad 
    \mathbb{E}(Y)=\mu+\sigma \frac{1-2\tau}{\tau(1-\tau)}.
\end{equation}

In a classical linear model, the response variable is assumed to be normally distributed and the focus is on modeling the location parameter (\textit{i.e.} the expectation) through a linear predictor. In parametric quantile regression, it is natural to assume that the response variable is distributed according to an $\mathcal{AL}$ distribution \citep{geraci_2014_lqmm}. 
Since the location parameter $\mu$ corresponds to the quantile of order $\tau$ of the distribution (see equation \eqref{eq:ALD_cum_distr}), classical quantile linear regression consists in fixing the parameter $\tau$ to the order of the quantile of interest (e.g. $0.25$ or $0.5$ for the first quartile and the median, respectively) and defining the location parameter $\mu$ by a linear predictor.
Furthermore, assuming the $\mathcal{AL}$ distribution and estimating the parameters of the quantile regression model using the maximum likelihood method is equivalent to the standard approach, which obtains parameter estimates by minimizing the sum of quantile loss functions \eqref{eq:rho} applied to model errors. 
For further details, we refer the reader to the works of \cite{yu_bayesian_2001} and \cite{geraci_2007}, among others.
Even if an exact equivalence does not hold anymore in mixed model framework, the link with quantile regression remains strong and fruitful.

In this work, the interest in this distribution is multiple for both inference and interpretation. Firstly, its three parameters give a flexible fit to the data, allowing to assess the impact of explanatory variables on location, scale and skewness. Unlike conventional approaches, the fit of skewness parameter is used to manage the asymmetry of the data around the location part. By fitting the skewness parameter, the location parameter corresponds to the mode of the distribution (see Figure \ref{fig_ADL_rho}) driven by the data, and is not computed as the quantile of interest. In this way, the location parameter corresponds to the central behavior of the distribution in the same way as the mean or median.

\subsection{Model specifications}

Let $\boldsymbol{Y_i}=(Y_{i1},\ldots,Y_{in_i})^{\top}$ denotes the vector of the $n_i$ measurements for the subject $i(i=1,\ldots,n)$. The response variable $Y_{ij}$ is assumed to be distributed from the $\mathcal{AL}$ distribution conditionally to subject-specific random effects $\textbf{r}_i$ such as
\begin{equation*}\label{eq:y_cond_dist}
Y_{ij}\vert \textbf{r}_i \; \sim \; \mathcal{AL}\left(\; \mu_{ij}, \; \sigma_{ij}, \; \tau_{ij} \;  \right) ,
\end{equation*}
where the parameters of location $\mu_{ij}$, scale $\sigma_{ij}$ and skewness $\tau_{ij}$ can be defined as a function of a linear predictor that includes fixed and random effects. 
In this section, we distinguish between three different models, depending on the assumptions about the scale and skewness parameters: the linear quantile mixed model (LQMM), the location-scale linear quantile mixed model (LSLQMM) and the asymmetric Laplace distribution regression model (ALDRM).

\subsubsection{Linear quantile mixed model}\label{sec:LQMM}

The LQMM is the classic quantile regression model for repeated measurements \citep{geraci_2014_lqmm}. For a quantile of interest of order $\tau\in ]0;1[$, it is defined by
\begin{eqnarray*}
Y_{ij} & = & \textbf{x}_{ij,\mu}^\top\boldsymbol{\beta} +  \textbf{z}_{ij,\mu}^\top \textbf{b}_{i} +  \epsilon_{ij}, 
\end{eqnarray*}
where $\boldsymbol{\beta}$ is the $p_\beta$-length vector of quantile-specific fixed effects associated with the design vector $x_{ij,\mu}$, $\textbf{b}_{i}\sim\mathcal{N}\left(0,\Sigma_b\right)$ is the subject-specific random effect vector associated with the design vector $\textbf{z}_{ij,\mu}$, $\epsilon_{ij}$ the residual error term assumed independent of $\textbf{b}_{i}$ and $\epsilon_{ij}\sim\mathcal{AL}\left(0,\sigma,\tau\right)$ in a parametric framework. 
Thus,
\begin{equation*}\label{eq:lqmm}
Y_{ij}\vert \textbf{b}_{i} \sim\mathcal{AL}\left(\mu_{ij},\sigma,\tau\right) 
\quad \text{with} \quad 
\mu_{ij}= \textbf{x}_{ij,\mu}^\top\boldsymbol{\beta} +  \textbf{z}_{ij,\mu}^\top \textbf{b}_{i},
\end{equation*}
where the scale $\sigma\in\mathbb{R}^+$ is an unknown parameter and the skewness parameter $\tau\in ]0,1[$ is previously fixed to the order of quantile of interest. Indeed, given equation \eqref{eq:ALD_cum_distr}, $\mu_{ij}$ corresponds to the quantile with order $\tau\in ]0,1[$ of the conditional distribution associated with the measurement $j$ of subject $i$.
\textcolor{black}{However, LQMM fitting is not appropriate when data exhibit heteroscedasticity, as it assumes a common scale parameter for all observations/residuals.}

\subsubsection{Location-scale linear quantile mixed model}\label{sec:LSLQMM}

\textcolor{black}{When data present heteroscedasticity, it is possible to take this into account by introducing an indexed parameter $\sigma_i$ for the scale parameter, which now depends on the individual. This has already been proposed in the work of \citet{alcaraz_lqmm_2025}, in which the authors propose a penalization approach by considering the $\sigma_i$ as fixed.
To make modeling more flexible and understandable, we first propose in this work} to relax the assumption of a fixed scale parameter for the whole population or for each individual by fitting it according to covariates and subject-specific random effects by using a logarithmic link function as in a LSMM \citep{courcoul_SIM_2025}. Therefore, we defined the LSLQMM by 
\begin{equation*}
Y_{ij}\vert \textbf{r}_i \sim \mathcal{AL}\left( \mu_{ij}, \sigma_{ij}, \tau  \right)
\quad \text{with} \quad
\left\lbrace 
\begin{array}{rcl}
\mu_{ij}\; & = & \textbf{x}_{ij,\mu}^\top\boldsymbol{\beta} \; + \;  \textbf{z}_{ij,\mu}^\top \textbf{b}_{i} \\
\log\left(\sigma_{ij}\right) & = & \textbf{x}_{ij,\sigma}^\top\boldsymbol{\xi} \; + \;  \textbf{z}_{ij,\sigma}^\top \textbf{u}_{i} 
\end{array}
\right.
\end{equation*}
where 
the location part is already defined in section \ref{sec:LQMM},
$\boldsymbol{\xi}$ is the $p_\sigma$-length vector of scale fixed effects associated with design vector $\textbf{x}_{ij,\sigma}$, $\textbf{u}_{i}\sim\mathcal{N}\left(\textbf{0},\Sigma_u\right)$ is the subject-specific random effect vector associated with design vector $\textbf{z}_{ij,\sigma}$. 
We denote by $\textbf{r}_i=\left(\textbf{b}_{i}^\top,\textbf{u}_{i}^\top\right)^\top$ the vector that includes all subject-specific random effects, and the random effect vectors $\textbf{b}_{i}$ and $\textbf{u}_{i}$ are assumed to be independent, as is classically done in linear regression between location and residual parts.

This model takes into account heteroskedasticity in the data by defining the scale parameter as a function of fixed effects and subject-specific random effects (for an example, please refer to \citet{alcaraz_2025_ecolo}). 
However, the skewness parameter is always a parameter previously fixed by user to the order of the quantile of interest. LSLQMM therefore does not handle the asymmetry present in the data, either at population or group level. At the population level, one solution for managing asymmetry is to select the value of $\tau$ using a model selection procedure. A better alternative is to estimate the $\tau$ parameter simultaneously with other parameters. However, assuming the same asymmetry for each individual distribution is a strong assumption. 

\subsubsection{Asymmetric Laplace distribution regression model}\label{sec:aldrm}

To relax the assumption of asymmetry common to the entire population, we extend the LSLQMM by fitting the skewness parameter according to covariates and subject-specific random effects by using the logit link function. Therefore, we defined the ALDRM by 
\begin{equation}\label{eq:ALDRM}
Y_{ij}\vert \textbf{r}_i \sim \mathcal{AL}\left( \mu_{ij}, \sigma_{ij}, \tau_{ij}  \right)
\quad \text{with} \quad
\left\lbrace 
\begin{array}{rcl}
\mu_{ij}\; & = & \textbf{x}_{ij,\mu}^\top\boldsymbol{\beta} +  \textbf{z}_{ij,\mu}^\top \textbf{b}_{i} \\
log\left(\sigma_{ij}\right) & = & \textbf{x}_{ij,\sigma}^\top\boldsymbol{\xi} + \textbf{z}_{ij,\sigma}^\top \textbf{u}_{i}  \\
logit\left(\tau_{ij}\right) & = & \textbf{x}_{ij,\tau}^\top\boldsymbol{\alpha} + \textbf{z}_{ij,\tau}^\top \textbf{a}_{i}
\end{array}
\right.
\end{equation}
where the location and scale parts are both defined in sections \ref{sec:LQMM} and \ref{sec:LSLQMM}, respectively, $\boldsymbol{\alpha}$ is the $p_\alpha$-length vector of skewness fixed effects associated with design vector $\textbf{x}_{ij,\tau}$, $\textbf{a}_{i}\sim\mathcal{N}\left(\textbf{0},\Sigma_a\right)$ is a subject-specific random effect vector associated with design vector $\textbf{z}_{ij,\tau}$, and $\textbf{r}_i=\left(\textbf{b}_{i}^\top,\textbf{u}_{i}^\top,\textbf{a}_{i}^\top\right)^\top$ the vector including all subject-specific random effects, and the random effect vectors $\textbf{b}_{i}$, $\textbf{u}_{i}$ and $\textbf{a}_{i}$ are assumed to be independent. 
By modeling the component related to skewness, this model fully aligns with a parametric distributional regression approach, in which all parameters are modeled using linear predictors through appropriate link functions.

\subsection{Estimation procedure}\label{sec:estimation}


In this work, we use the Bayesian framework. Instead of directly considering the $\mathcal{AL}$ distribution, we use the Gaussian mixture representation which is often used for quantile regression model estimation \citep{kozumi_2011,kotz_laplace_2012,waldmann_2015,petrella_2019,yang_2019}. For $Y \sim\mathcal{AL}\left( \mu,\sigma,\tau \right)$, the rewriting of the $\mathcal{AL}$ distribution is derived from the following equation
\begin{equation*}\label{eq:al_to_NE}
f_Y\left(y \right) = \int_{\mathbb{R}^+}f_{Y\vert W}\left(y\vert W=w \right)f_{W}\left(w \right)dw 
\end{equation*}
with
\begin{eqnarray*}
\left\{ \begin{array}{rcl}
Y\vert \mu,\sigma,\tau, W=w & \sim & \mathcal{N}\Big( \mu+c_1\left(\tau\right) w,\; c_2\left(\tau\right)\sigma w \Big) \\
W\vert \sigma & \sim & \mathcal{E}xp \left(\frac{1}{\sigma} \right)
\end{array} 
\right. ,
\end{eqnarray*}
where $c_1\left(\tau\right)=\frac{1-2\tau}{\tau\left(1-\tau\right)}$, $c_2\left(\tau\right)=\frac{2}{\tau\left(1-\tau\right)}$ and $W$ is an auxiliary random variable exponentially distributed such as $\mathbb{E}\left(W\right)=\sigma$. \\

Considering a Bayesian framework, the ALDRM is defined as a hierarchical Bayesian model using the Gaussian mixture representation of the $\mathcal{AL}$ distribution  
\begin{eqnarray}\label{eq:distributions}
Y_{ij}\vert W_{ij}=w_{ij}, \textbf{r}_{i}, \boldsymbol{\theta}  & \sim & \mathcal{N}\Big( \mu_{ij} + c_1\left(\tau_{ij}\right) w_{ij},\; c_2\left(\tau_{ij}\right)\sigma_{ij} w_{ij} \Big) \nonumber\\
W_{ij}\vert \boldsymbol{\xi}, \textbf{u}_{i}  & \sim & \mathcal{E}xp \left(\frac{1}{\sigma_{ij}} \right) \quad \text{with }\; \mathbb{E}\left(W_{ij}\right)= \sigma_{ij} \; , \nonumber\\
\textbf{b}_{i}\vert \Sigma_b & \sim & \mathcal{N}\Big( \textbf{0},\; \Sigma_b \Big) \\
\textbf{u}_{i}\vert \Sigma_u & \sim & \mathcal{N}\Big( \textbf{0},\; \Sigma_u \Big) \nonumber\\
\textbf{a}_{i}\vert \Sigma_a & \sim & \mathcal{N}\Big( \textbf{0},\; \Sigma_a \Big) \nonumber  
\end{eqnarray}
where $\boldsymbol{\theta} =\left( \boldsymbol{\beta}^\top, \Sigma_b, \boldsymbol{\xi}^\top, \Sigma_u, \boldsymbol{\alpha}^\top, \Sigma_a \right)^\top$ is the vector of model parameters, and $\mu_{ij}$, $\sigma_{ij}$ et $\tau_{ij}$ are defined in equation \eqref{eq:ALDRM}.

The posterior distribution of parameters is:
\begin{equation*}\label{eq:posterior_distrib}
\pi\left(\boldsymbol{\theta},\boldsymbol{r},\boldsymbol{w}\vert \boldsymbol{y}\right) \propto \prod^{n}_{i=1} \prod^{n_i}_{j=1} f\left( y_{ij} \vert w_{ij}, \textbf{r}_{i}, \boldsymbol{\theta} \right)f\left(w_{ij}\vert\boldsymbol{\theta}\right) \phi\left( \textbf{r}_{i}\vert\Sigma_r\right) \pi\left(\boldsymbol{\theta}\right) 
\end{equation*}
where $f\left(.\right)$ are density functions associated with distributions defined in \eqref{eq:distributions}, $\phi$ denotes the zero-mean normal density function for subject-specific random effects and $\pi\left(\boldsymbol{\theta}\right)$ is the prior distribution of parameters that are assumed to be independent.  
Concerning prior distributions of parameters, we considered normal distributions for regression parameters and inverse Wishart ($\mathcal{IW}$) distributions for the covariance matrices of random effects. For both LSLQMM and ALDRM, the following priors are considered: 
\begin{equation*}
\boldsymbol{\beta} \sim \mathcal{N}_{p_\beta} \Big( \boldsymbol{\mu}_\beta , \; \Sigma_\beta \Big), 
\quad \boldsymbol{\xi} \sim \mathcal{N}_{p_\sigma} \Big( \boldsymbol{\mu}_\xi , \; \Sigma_\xi \Big), 
\quad \Sigma_b \sim \mathcal{IW}\Big(\boldsymbol{\omega}_b, \; \Omega_b \Big) \text{ and }
\Sigma_u \sim \mathcal{IW}\Big(\boldsymbol{\omega}_u, \; \Omega_u \Big),
\end{equation*}
and the additional priors of skewness-part parameters in ALDRM are
$$
\boldsymbol{\alpha} \sim \mathcal{N}_{p_\alpha} \Big( \boldsymbol{\mu}_\alpha , \; \Sigma_\alpha \Big)
\text{ and }
\Sigma_a \sim \mathcal{IW}\Big(\boldsymbol{\omega}_a, \; \Omega_a \Big).
$$
These priors can be considered as vague or data-driven using simpler previous estimated models. In this work, we considered them as vague. \\

The estimation of models is implemented in the \texttt{R}-package \texttt{BeQut} and statistical inference is based on posterior parameter samples. Bayesian Markov chain Monte Carlo (MCMC) sampling is performed using the \texttt{JAGS} software \citep{plummer_2016}. The open source \texttt{R}-package \texttt{BeQut} is available on Github at the following link: \url{https://github.com/AntoineBbi/BeQut}.

\subsection{Model selection strategy}\label{sec:criteria}

In the context of distributional regression, the choice of model relates to the probability distribution and the predictors, as well as to the link functions that connect the distribution parameters to the predictors \citep{GAMLSS_2024}. 
The comparison of model fits is done classically using information criteria dealing with a compromise between fit and complexity. 
The definition of these criteria depends on the method used to estimate the parameters (frequentist or Bayesian), and when the regression model includes random effects, the choice of information criterion may also depend on whether the conditional or marginal version is used. 
In the Bayesian approach, the Deviance Information Criterion (DIC) is commonly used in practice, but it has drawbacks. In particular, this criterion is not invariant to model reparameterization \citep{2018_Quintero, GAMLSS_2024}. Since the estimation procedure of the ALDRM is based on a rewriting of the $\mathcal{AL}$ distribution, it is not suitable for comparison with another distribution. 

Another strategy would be to study the predictive capabilities of models based on mean square error (MSE) or mean absolute error (MAE). However, these two criteria are highly dependent on the choice of the prediction. In fact, the MSE will favor the LSMM (see its specification in appendix \ref{sec:lsmm}), which assumes a Gaussian response variable, because the location parameter corresponds to the mean, while the MAE will favor the LSLQMM when $\tau$ is set to 0.5 because the location parameter corresponds to the median. Since the location parameter of ALDRM corresponds to the mode, it will be penalized relative to LSMM and LSLQMM. To be fair, the estimation of the mean (see equation \eqref{eq:var_exp}) or median (see equation \eqref{eq:ALD_quantile_fct}) should be considered for ALDRM. 
Second, these criteria only evaluate the adequacy of the central behavior, whereas the objective of the proposed distributional regression models is the fit of the entire distribution of the response variable. \\

In this work, we are interested in evaluating the adequacy of the estimated distribution for the data of interest. To this end, we can consider using indices derived from the Quantile-Quantile-plot (QQ-plot), such as the (root) mean square errors between the empirical and theoretical quantiles. 
This criterion can therefore be used on grouped data provided that there is no temporal dependency. 
However, when adjusting longitudinal data where only one measurement is available at each time point and the distribution of the variable is assumed to vary over time, a criterion derived from the QQ-plot is not applicable. 
Assuming that there is an individual distribution that varies over time, we propose a new measure of the adequacy of the data to the adjusted parametric distribution. 
Let $\Gamma$ be the set of orders $\gamma\in]0,1[$ of the quantiles on which the criterion will be constructed. For the individual $i=1,\ldots,n$, we define the divergence between the adjusted individual distribution and the empirical distribution by
\begin{equation}\label{eq:criterCi}
    C_{i,\Gamma} = \frac{1}{\#\Gamma}\sum_{\gamma\in \Gamma} \ell\left( \widehat{\gamma}_i-\gamma \right)
\end{equation}
where $\ell(.)$ a loss function and $\widehat{\gamma}_i=\frac{n_i^\gamma}{n_i}$ with $n_i^\gamma=\sum_{j=1}^{n_i}\mathds{1}\left\{y_{ij}<\widehat{\mathcal{Q}}_{Y_{ij}\vert \textbf{r}_i}(\gamma)\right\}$ the number of observations below the quantile of order $\gamma$ from the individual distribution predicted by the model for the measurement $j$ of individual $i$, denoted by $\widehat{\mathcal{Q}}_{Y_{ij}\vert \textbf{r}_i}(\gamma)$. Considering the trajectory of the $\gamma$th quantile over time derived from the estimated distribution, $\widehat{\gamma}_i$ represents the proportion of measurements below this trajectory (see Figure \ref{fig:id_points_mode_quantile} in the application section). 
Thus, among several competing distributional regression mixed models, the best one to fit the individual distribution of the subject $i$ is the one that minimizes this criterion. For all data, the best model is the one that minimizes 
\begin{equation}\label{eq:criterC}
C_{\Gamma}=\frac{1}{n}\sum_{i=1}^n C_{i,\Gamma}.
\end{equation}
Note that the criterion we proposed in equation \eqref{eq:criterCi} is based on a loss function applied to an error. Taking into account the quadratic cost function or the absolute cost function, $C_{\Gamma}$ corresponds to the mean of mean square error (MMSE) or the mean of mean absolute error (MMAE), respectively.

\section{Simulation study}

\subsection{Design}

The aim of the Monte Carlo simulation study is first to numerically validate the procedure of estimation, and then to study the goodness-of-fit criteria for model selection. 
Using posterior means as parameter estimates, the estimation procedure is evaluated using average bias and the coverage rate (CR) for the 95\% credibility interval of the parameter.
The simulation study also investigates the sensitivity of the estimates to the number of subjects $n$ and the number of repeated measurements $n_i$ per subject. 

Then, we compare the fits of the ALDRM, of the LSLQMM with $\tau=0.5$ (implying that the location parameter is the median), and of the LSMM using a Gaussian assumption. The LSMM and its estimation procedure are presented in appendix \ref{sec:lsmm}.
To do that, we consider the criteria proposed in Section \ref{sec:criteria} based on MMSE and MMAE with three sets of orders: Set 1 denoted by $\Gamma_1=\left\{0.1,0.2,\ldots,0.9\right\}$ including all deciles, Set 2 denoted by $\Gamma_2=\left\{0.25,0.5,0.75\right\}$ including all quartiles, and Set 3 denoted by $\Gamma_3=\left\{0.1,0.5,0.9\right\}$ including median, first and ninth deciles. 
When the adjustment focuses on a current value of the biomarker, the three models are also compared using the $MSE=\frac{1}{\sum_{i=1}^n n_i} \sum_{i=1}^n\sum_{j=1}^{n_i} (y_{ij}-\widehat{y}_{ij})^2$ and $MAE=\frac{1}{\sum_{i=1}^n n_i} \sum_{i=1}^n\sum_{j=1}^{n_i} \vert y_{ij}-\widehat{y}_{ij}\vert$ criteria. The predicted current value is
$$
\widehat{y}_{ij}= \left\{
\begin{array}{l}
    \widehat{\mu}_{ij} \quad\text{with}\quad Y_{ij}|\textbf{r}_i\sim\mathcal{N}({\mu}_{ij},{\sigma}_{ij}) \quad\text{for the LSMM} \\
     \widehat{\mu}_{ij} \quad \text{with}\quad Y_{ij}|\textbf{r}_i\sim\mathcal{AL}({\mu}_{ij},{\sigma}_{ij},\tau=0.5)  \quad\text{for the LSLQMM} 
\end{array}
\right.
$$
For the LSMM, the current value predicted by the model is the conditional expectation, but also the median and mode for the assumed Gaussian distribution. For the LSLQMM with $\tau=0.5$, the current value predicted by the model is the median, but also the expectation (see equation \eqref{eq:var_exp}) and the mode for the assumed $\mathcal{AL}$ distribution. 
For the ALDRM, the location parameter is the mode. However, given that the MSE and MAE criteria favor predictions of the expectation and median, respectively, these values for the ALDRM will also be considered in the comparison and derived from equations \eqref{eq:var_exp} and \eqref{eq:ALD_quantile_fct} for $p=0.5$. \\

Data generation is based on the ALDRM defined in equation \eqref{eq:ALDRM} in which both the three linear predictors and parameter values used are driven by the application results (cf. section \ref{sec:application}). 
The generation procedure is as follows:
\begin{enumerate}
\item Fix the value of parameters $\boldsymbol{\beta}$, $\boldsymbol{\xi}$, $\boldsymbol{\alpha}$, $\Sigma_b$, $\Sigma_u$ and $\Sigma_a$;
\item Draw subject-specific random effects for $i=1,\ldots,n$ such as
\begin{equation*}
\textbf{r}_i =
\begin{pmatrix}
    \textbf{b}_i \\
    \textbf{u}_i \\
    \textbf{a}_i \\
\end{pmatrix}
\sim \mathcal{N} \left(
\begin{pmatrix} 
    \textbf{0}  \\
    \textbf{0}  \\
    \textbf{0}  \\
\end{pmatrix}
\; , \; 
\begin{pmatrix} 
    \Sigma_b & 0 & 0 \\
    0 & \Sigma_{u} & 0 \\
    0 & 0 & \Sigma_{a} \\
\end{pmatrix}
\right).
\end{equation*}
\item Draw 2 subject-specific covariates, $X_{i1}\sim\mathcal{N}\left(0,1\right)$ and $X_{i2}\sim\mathcal{B}er\left(0.5\right)$;
\item Define $n_i=m$ measurements equally spaced between $0$ and $t_{max}=10$;
\item Generate the response variable using the function \texttt{rALD} of package \texttt{ald} \citep{ald_2015} such as $Y_{ij}\vert \textbf{r}_i \sim \mathcal{AL}\left( \mu_{ij}, \sigma_{ij}, \tau_{ij}  \right)$ for the measurement $j=1,\ldots,m$ of individual $i$ where 
\begin{equation}\label{eq:ALDRM_sim}
\left\{
\begin{array}{rcl}
\mu_{ij}\; & = & \beta_0 + \beta_1 t_{ij} + \beta_2 t_{ij}^2 + \beta_3 X_{i1} + \beta_4 X_{i2}  + b_{i0} + b_{i1} t_{ij} + b_{i2} t_{ij}^2 \\
log\left(\sigma_{ij}\right) & = & \xi_0 + \xi_1 t_{ij} + \xi_2 X_{i1}  + u_{i0} + u_{i1} t_{ij} \\
logit\left(\tau_{ij}\right) & = & \alpha_1 X_{i1} + \alpha_2 X_{i2} + a_{i0} + a_{i1} t_{ij} \\
\end{array}
\right. \; 
\end{equation}
\end{enumerate}
The values used for the parameters (step 1), the covariates (step 3) and the model in equation \eqref{eq:ALDRM_sim} (step 5) are inspired by the application results (see true parameter values in Tables \ref{tab:simu_n} and \ref{tab:simu_m}). 
The default scenario considered in this simulation study includes $n=200$ subjects with the same number of repeated measurements $m=50$ for all subjects. 
To evaluate the influence of the number of subjects and the number of repeated measurements, we consider different values such as $n\in\{200,500,1000\}$ and $m\in\{10,20,50,200\}$, respectively.

\subsection{Results}

Table \ref{tab:simu_n} shows that the estimations are unbiased with a good coverage rate of the credibility interval, except for the variance parameter of the random slope associated with the skewness component, whatever the number of subjects considered $n$. However, as the number of subjects increases, the parameter estimation and coverage rates improve. 

\begin{table}[t!]
\caption{Simulation results with different numbers of subject $n$, obtained on $N=500$ datasets generated from the {ALDRM} with $m=50$ repeated measurements. $\overline{\widehat{\theta}}$, $\overline{\widehat{sd\left(\theta\right)}}$ are the average of posterior means and the standard deviations, respectively. CR is the coverage rate corresponding to the percentage of times the true parameter $\theta$ falls within the 95\% credibility interval.}\label{tab:simu_n}
\scalebox{0.9}{
\begin{tabular}{cccccccccccccc}
\hline 
\multicolumn{2}{c}{} && \multicolumn{3}{c}{\texttt{\textbf{$n=200$}}} && \multicolumn{3}{c}{\texttt{\textbf{$n=500^a$}}} && \multicolumn{3}{c}{\texttt{\textbf{$n=1000^b$}}} \\ 
\cline{4-6}\cline{8-10}\cline{12-14} \\
 \multicolumn{2}{l}{} & $\theta$ 
 &  $\overline{\widehat{\theta}}$ & $\overline{\widehat{sd\left(\theta\right)}}$ & CR 
 && $\overline{\widehat{\theta}}$ & $\overline{\widehat{sd\left(\theta\right)}}$ & CR 
 && $\overline{\widehat{\theta}}$ & $\overline{\widehat{sd\left(\theta\right)}}$ & CR \\ 
\hline
\multicolumn{3}{l}{\textbf{Location part}} &   &   &  &   \\  
  & \multicolumn{1}{l}{$\beta_0$}  &  13.000 
  & 12.946 & 0.176 & 0.954 &  & 12.985 & 0.111 & 0.941 &  & 12.991 & 0.079 & 0.937 \\  
  & \multicolumn{1}{l}{$\beta_1$}  &   0.300 
  & 0.302 & 0.043 & 0.956 &  & 0.299 & 0.027 & 0.945 &  & 0.302 & 0.019 & 0.933 \\ 
  & \multicolumn{1}{l}{$\beta_2$}  &  -0.030 
  & -0.030 & 0.008 & 0.962 &  & -0.030 & 0.005 & 0.939 &  & -0.030 & 0.003 & 0.944 \\
  & \multicolumn{1}{l}{$\beta_3$}  &   0.600 
  & 0.609 & 0.121 & 0.938 &  & 0.599 & 0.076 & 0.953 &  & 0.596 & 0.054 & 0.946 \\ 
  & \multicolumn{1}{l}{$\beta_4$}  &   0.800 
  & 0.845 & 0.238 & 0.932 &  & 0.810 & 0.151 & 0.943 &  & 0.806 & 0.107 & 0.956 \\ 
  & \multicolumn{1}{l}{$\Sigma_{b,11}$}  & 3.000 
  & 3.017 & 0.344 & 0.960 &  & 3.016 & 0.216 & 0.943 &  & 3.010 & 0.152 & 0.946 \\ 
  & \multicolumn{1}{l}{$\Sigma_{b,22}$}  &  0.300 
  & 0.299 & 0.037 & 0.958 &  & 0.299 & 0.023 & 0.953 &  & 0.300 & 0.016 & 0.958 \\  
  & \multicolumn{1}{l}{$\Sigma_{b,33}$}  &  0.010 
  & 0.011 & 0.001 & 0.950 &  & 0.010 & 0.001 & 0.961 &  & 0.010 & 0.000 & 0.950 \\  
  & \multicolumn{1}{l}{$\Sigma_{b,12}$}  & -0.320 
  & -0.315 & 0.086 & 0.958 &  & -0.322 & 0.054 & 0.935 &  & -0.321 & 0.038 & 0.954 \\ 
  & \multicolumn{1}{l}{$\Sigma_{b,13}$}  &  0.014 
  & 0.014 & 0.014 & 0.962 &  & 0.014 & 0.009 & 0.963 &  & 0.014 & 0.006 & 0.950 \\  
  & \multicolumn{1}{l}{$\Sigma_{b,23}$}  &  -0.025 
  & -0.025 & 0.005 & 0.940 &  & -0.025 & 0.003 & 0.957 &  & -0.025 & 0.002 & 0.937 \\   
\multicolumn{3}{l}{\textbf{Scale part}} &   &   &  \\ 
  & \multicolumn{1}{l}{$\xi_0$}  &  {-0.600} 
  & -0.597 & 0.035 & 0.952 &  & -0.599 & 0.021 & 0.953 &  & -0.600 & 0.016 & 0.942 \\ 
  & \multicolumn{1}{l}{$\xi_1$}  &  {-0.070} 
  & -0.069 & 0.015 & 0.946 &  & -0.070 & 0.007 & 0.947 &  & -0.070 & 0.005 & 0.931 \\   
  & \multicolumn{1}{l}{$\xi_2$}  &   0.084 
  & 0.083 & 0.026 & 0.938 &  & 0.083 & 0.016 & 0.918 &  & 0.085 & 0.012 & 0.942 \\  
  & \multicolumn{1}{l}{$\Sigma_{u,11}$}  & 0.060 
  &  0.060 & 0.019 & 0.956 &  & 0.059 & 0.011 & 0.937 &  & 0.060 & 0.009 & 0.948 \\   
  & \multicolumn{1}{l}{$\Sigma_{u,22}$}  & {0.010} 
  &  0.013 & 0.006 & 0.936 &  & 0.010 & 0.001 & 0.939 &  & 0.010 & 0.001 & 0.956 \\ 
  & \multicolumn{1}{l}{$\Sigma_{u,12}$}  & -0.003 
  &  -0.002 & 0.007 & 0.968 &  & -0.003 & 0.003 & 0.957 &  & -0.003 & 0.002 & 0.962 \\  
\multicolumn{3}{l}{\textbf{Skewness part}} &   &   &  \\ 
  & \multicolumn{1}{l}{$\alpha_0$}  &  0.130 
  &  0.132 & 0.052 & 0.968 &  & 0.130 & 0.033 & 0.929 &  & 0.131 & 0.023 & 0.929 \\ 
  & \multicolumn{1}{l}{$\alpha_1$}  &  0.150 
  &  0.150 & 0.074 & 0.926 &  & 0.149 & 0.046 & 0.945 &  & 0.151 & 0.033 & 0.912 \\ 
  & \multicolumn{1}{l}{$\Sigma_{a,11}$}  & 0.250 
  &  0.262 & 0.058 & 0.950 &  & 0.254 & 0.037 & 0.967 &  & 0.252 & 0.027 & 0.958 \\  
  & \multicolumn{1}{l}{$\Sigma_{a,22}$}  & {0.050} 
  &  0.057 & 0.007 & 0.884 &  & 0.053 & 0.004 & 0.916 &  & 0.051 & 0.003 & 0.927 \\   
  & \multicolumn{1}{l}{$\Sigma_{a,12}$}  & -0.020 
  & -0.022 & 0.014 & 0.966 &  & -0.021 & 0.009 & 0.984 &  & -0.020 & 0.006 & 0.948 \\
\hline
\multicolumn{8}{l}{$^a$ 490 samples with satisfying convergence; } \\
\multicolumn{8}{l}{$^b$ 479 samples with satisfying convergence; } \\
\end{tabular} 
}
\end{table}  

Table \ref{tab:simu_m} allows us to study the influence of the number of repeated measurements on the estimation of the model. The estimation of parameters is not affected in the presence of intensive data ($m=200$), with similar results when this number $m$ varies between 10, 20 (see Table \ref{tab:simu_n} for $m=50$) and 200 measurements per individual. 
Slight estimation biases occur when $m=10$, especially for variance parameters associated with random effects, and therefore low coverage rates appear. 
Overall, the results show that the estimation procedure is robust and remains accurate for both high and low values of the number of subjects and the number of repeated measurements per subject. 

\begin{table}[t!]
\caption{Simulation results with different numbers of repeated measurements $m$, obtained on $N=500$ datasets generated from the {ALDRM} with $n=200$ subjects. $\overline{\widehat{\theta}}$, $\overline{\widehat{sd\left(\theta\right)}}$ are the average of posterior means and the standard deviations, respectively. CR is the coverage rate corresponding to the percentage of times the true parameter $\theta$ falls within the 95\% credibility interval.}\label{tab:simu_m}
\scalebox{0.9}{
\begin{tabular}{cccccccccccccc}
\hline 
\multicolumn{2}{c}{} && \multicolumn{3}{c}{\texttt{\textbf{$m=10$}}} && \multicolumn{3}{c}{\texttt{\textbf{$m=20$}}} && \multicolumn{3}{c}{\texttt{\textbf{$m=200^*$}}} \\ 
\cline{4-6}\cline{8-10}\cline{12-14} \\
 \multicolumn{2}{l}{} & $\theta$ 
 &  $\overline{\widehat{\theta}}$ & $\overline{\widehat{sd\left(\theta\right)}}$ & CR 
 && $\overline{\widehat{\theta}}$ & $\overline{\widehat{sd\left(\theta\right)}}$ & CR 
 && $\overline{\widehat{\theta}}$ & $\overline{\widehat{sd\left(\theta\right)}}$ & CR \\ 
\hline
\multicolumn{3}{l}{\textbf{Location part}} &   &   &  &   \\  
  & \multicolumn{1}{l}{$\beta_0$}  &  13.000 
  &  12.954 & 0.197 & 0.958 & 
  & 12.953 & 0.185 & 0.944 & 
  & 12.960 & 0.170 & 0.941 \\
  & \multicolumn{1}{l}{$\beta_1$}  &   0.300 
  & 0.307 & 0.054 & 0.938 & 
  & 0.307 & 0.048 & 0.952 & 
  & 0.303 & 0.040 & 0.950 \\ 
  & \multicolumn{1}{l}{$\beta_2$}  &  -0.030 
  &  -0.030 & 0.009 & 0.978 & 
  & -0.030 & 0.008 & 0.950 & 
  & -0.030 & 0.007 & 0.961 \\ 
  & \multicolumn{1}{l}{$\beta_3$}  &   0.600 
  &  0.591 & 0.141 & 0.944 & 
  & 0.600 & 0.129 & 0.960 & 
  & 0.597 & 0.117 & 0.950 \\
  & \multicolumn{1}{l}{$\beta_4$}  &   0.800 
  &  0.830 & 0.269 & 0.940 & 
  & 0.831 & 0.249 & 0.952 & 
  & 0.831 & 0.231 & 0.957 \\ 
  & \multicolumn{1}{l}{$\Sigma_{b,11}$}  & 3.000 
  &  2.916 & 0.462 & 0.930 & 
  & 2.973 & 0.390 & 0.962 & 
  & 2.997 & 0.314 & 0.959 \\ 
  & \multicolumn{1}{l}{$\Sigma_{b,22}$}  &  0.300 
  &  0.288 & 0.057 & 0.938 & 
  & 0.295 & 0.045 & 0.934 & 
  & 0.303 & 0.033 & 0.952 \\ 
  & \multicolumn{1}{l}{$\Sigma_{b,33}$}  &  0.010 
  &  0.011 & 0.001 & 0.958 & 
  & 0.011 & 0.001 & 0.936 & 
  & 0.011 & 0.001 & 0.943 \\ 
  & \multicolumn{1}{l}{$\Sigma_{b,12}$}  & -0.320 
  &  -0.286 & 0.130 & 0.912 & 
  & -0.307 & 0.103 & 0.934 & 
  & -0.321 & 0.076 & 0.941 \\ 
  & \multicolumn{1}{l}{$\Sigma_{b,13}$}  &  0.014 
  &  0.011 & 0.018 & 0.930 & 
  & 0.013 & 0.016 & 0.940 & 
  & 0.014 & 0.013 & 0.957 \\ 
  & \multicolumn{1}{l}{$\Sigma_{b,23}$}  &  -0.025 
  &  -0.024 & 0.007 & 0.938 & 
  & -0.025 & 0.006 & 0.948 & 
  & -0.025 & 0.005 & 0.961 \\ 
\multicolumn{3}{l}{\textbf{Scale part}} &   &   &  \\ 
  & \multicolumn{1}{l}{$\xi_0$}  &  {-0.600} 
  &  -0.600 & 0.067 & 0.960 & 
  & -0.602 & 0.047 & 0.958 & 
  & -0.599 & 0.023 & 0.938 \\ 
  & \multicolumn{1}{l}{$\xi_1$}  &  {-0.070} 
  & -0.079 & 0.018 & 0.926 & 
  & -0.074 & 0.013 & 0.934 & 
  & -0.070 & 0.009 & 0.952 \\ 
  & \multicolumn{1}{l}{$\xi_2$}  &   0.084 
  & 0.081 & 0.044 & 0.948 &  & 0.081 & 0.034 & 0.950 &  & 0.083 & 0.020 & 0.943 \\ 
  & \multicolumn{1}{l}{$\Sigma_{u,11}$}  & 0.060 
  &  0.064 & 0.037 & 0.994 &  & 0.058 & 0.026 & 0.982 &  & 0.061 & 0.010 & 0.957 \\ 
  & \multicolumn{1}{l}{$\Sigma_{u,22}$}  & {0.010} 
  &  0.012 & 0.003 & 0.976 &  & 0.011 & 0.002 & 0.968 &  & 0.011 & 0.001 & 0.904 \\ 
  & \multicolumn{1}{l}{$\Sigma_{u,12}$}  & -0.003 
  &  -0.006 & 0.008 & 0.998 &  & -0.004 & 0.006 & 0.998 &  & -0.003 & 0.003 & 0.941 \\ 
\multicolumn{3}{l}{\textbf{Skewness part}} &   &   &  \\ 
  & \multicolumn{1}{l}{$\alpha_0$}  &  0.130 
  &  0.126 & 0.092 & 0.948 &  & 0.131 & 0.070 & 0.930 &  & 0.130 & 0.041 & 0.908 \\ 
  & \multicolumn{1}{l}{$\alpha_1$}  &  0.150 
  &  0.140 & 0.131 & 0.948 &  & 0.146 & 0.099 & 0.950 &  & 0.151 & 0.056 & 0.897 \\ 
  & \multicolumn{1}{l}{$\Sigma_{a,11}$}  & 0.250 
  &  0.319 & 0.128 & 0.978 &  & 0.287 & 0.093 & 0.986 &  & 0.258 & 0.036 & 0.963 \\ 
  & \multicolumn{1}{l}{$\Sigma_{a,22}$}  & {0.050} 
  &  0.062 & 0.010 & 0.842 &  & 0.060 & 0.009 & 0.814 &  & 0.055 & 0.006 & 0.879 \\ 
  & \multicolumn{1}{l}{$\Sigma_{a,12}$}  & -0.020 
  &  0.027 & 0.026 & 0.996 &  & -0.025 & 0.020 & 0.992 &  & -0.021 & 0.010 & 0.975 \\ 
\hline
\multicolumn{10}{l}{$^*$ 437 samples with satisfying convergence; }
\end{tabular} 
}
\end{table}  

\begin{figure}[t]
\begin{center}
\includegraphics[scale=0.8]{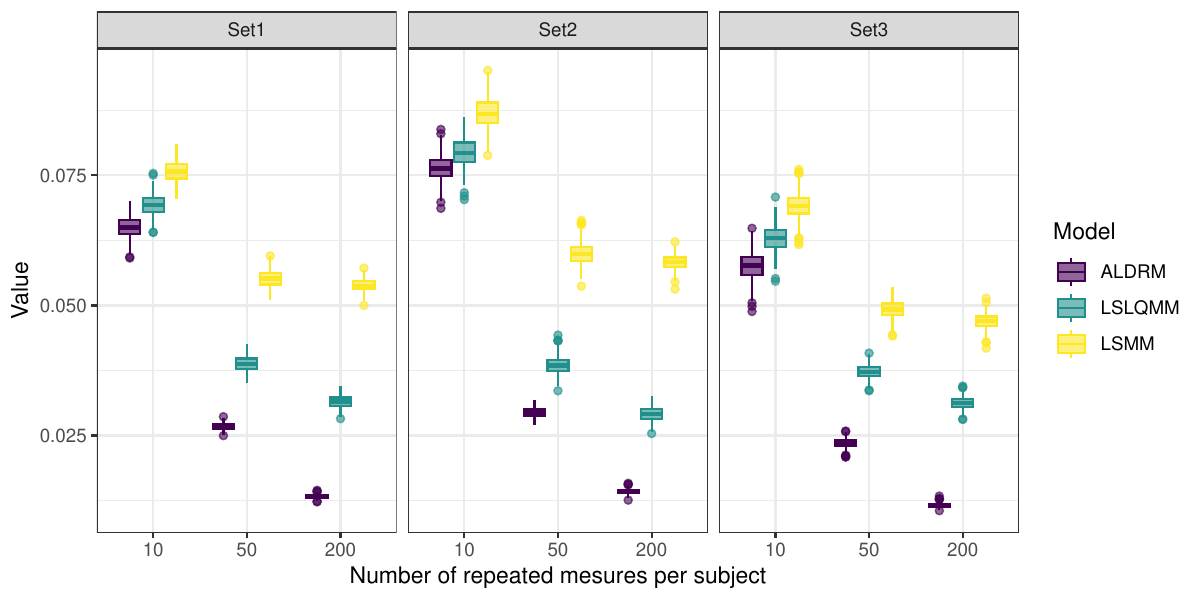}
\end{center}
\caption{Distributions of the $N=500$ computed criteria $C_\Gamma$ proposed in equation \eqref{eq:criterC} based on absolute cost loss function $\ell\left(x\right)=|x|$ for all $x\in\mathbb{R}$, the MMAE. Scenarii with $n=200$ and $m=10,50,200$ are considered. Set 1 denoted by $\Gamma_1=\left\{0.1,0.2,\ldots,0.9\right\}$ including all decile orders, Set 2 denoted by $\Gamma_2=\left\{0.25,0.5,0.75\right\}$ including all quartile orders, and Set 3 denoted by $\Gamma_3=\left\{0.1,0.5,0.9\right\}$ including median, first and ninth deciles.}\label{fig:boxplot_criteria_MMAE}
\end{figure}

Figure \ref{fig:boxplot_criteria_MMAE} shows the distributions of the selection criterion MMAE we proposed in equation \eqref{eq:criterC} according to the three models considered and the three sets of quantiles on which the criterion is calculated.
For the three sets considered, the simulation results show that the proposed criterion allows the correct model to be selected regardless of the number of repeated measurements considered $m$. However, we observe that the values of the criterion are closer between the different models considered when $m$ is low, suggesting that discrimination is more difficult. 
More precisely, regarding the percentage of model selection among the 500 replicates, ALDRM is always selected when $m=50,200$, whatever the set considered. When $m=10$, ALDRM is selected in 99\%, 86\%, and 98\% of cases for sets 1, 2, and 3, respectively. In other cases, LSLQMM is always chosen, whereas LSMM is never chosen.
Therefore, the $\mathcal{AL}$ distribution on which the generation model is based is always selected.
\textcolor{black}{Similar results were obtained for the MMSE (see Figure \ref{fig:boxplot_criteria_MMSE} in the appendix)}.

In cases where the goodness-of-fit is based on its ability to predict the observed value, Figure \ref{fig:boxplot_errors_m} in appendix shows that the choice of predicted value is important depending on whether MSE or MAE is considered. Indeed, if the value predicted by the three models corresponds to the prediction of the location parameter given random effects, then the MSE tends to favor the LSMM where the location parameter corresponds to the expectation, while the MAE tends to favor the LSLQMM where the location parameter corresponds to the median. Although the ALDRM is well specified, its location parameter corresponding to the mode is not appropriate to minimize these two criteria. On the other hand, when the predicted value by ALDRM is defined by the expectation or by the median (obtained directly by equations \eqref{eq:var_exp} and \eqref{eq:ALD_quantile_fct}, respectively), then the ALDRM is better in average than LSMM and LSLQMM in terms of MSE and MAE, respectively. However, it is not easy to distinguish the correct model from the others.
Note that the number of repeated measurements does not affect these results. 

\section{Application}\label{sec:application}

\subsection{Data and clinical motivations}\label{sec:app_data}

Data were collected from 201 patients hospitalized in intensive care unit of the Bordeaux University Hospital for subarachnoid hemorrhage (SAH) between June 2018 and June 2019. 
The patients were monitored during their stay, allowing automatic data collection every hour up to a maximum of 14 days. Consequently, longitudinal variables were collected intensively. For systolic blood pressure, we have 42,450 observations with a median of 224 repeated measurements per patient, a minimum of 10, and a maximum of 322.
In addition, demographic and clinical variables were collected at inclusion, such as age and sex of the patient and smoking status.

\begin{figure}[t]
\begin{center}
\includegraphics[scale=0.8]{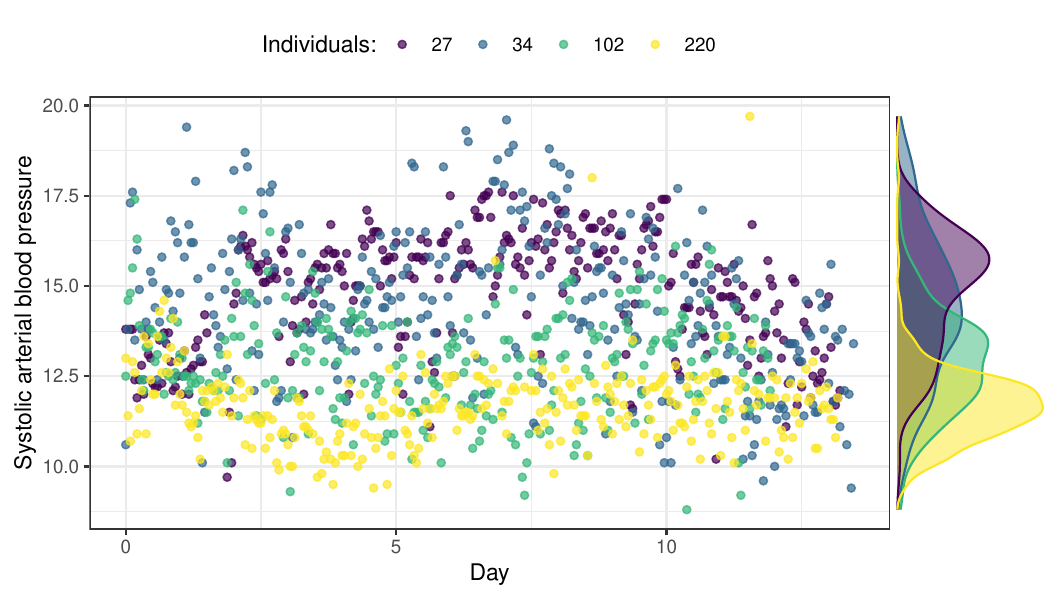}
\end{center}
\caption{Longitudinal measurements and empirical individual density for four selected individuals.}\label{fig:ID_scatterplot_density}
\end{figure}

In this application, we focus on fitting the systolic blood pressure over time. 
In this context, blood pressure is a biomarker of great interest because it is easily measurable and modifiable and also because many risk factors derive from it. 
Given that high blood pressure and blood pressure variability are known to be risk factors for stroke and other cerebrovascular and cardiovascular events \citep{pringle_2003,De_Courson_2021, courcoul_SIM_2025}, it could also be an indicator of poor prognosis and risk of complications among patients with SAH. Thus flexible modeling of its evolution post SAH and associated factors is of major clinical interest.
A good understanding of the evolution of the distribution of these measurements and the impact of covariates on the distribution of blood pressure would provide additional information to help in the management of patients.
In a neuroreanimation unit, although blood pressure is monitored and adjusted if necessary with medication to ensure a sufficient penetration force into the organs, it is highly heterogeneous between patients.
For example, Figure \ref{fig:ID_scatterplot_density} shows the measurements of four patients over time. We can observe different blood pressure levels (e.g. patient 220 has a lower level than patient 27), different variabilities (e.g. the variability of patient 220's measurements is smaller than that of patient 34), outliers (e.g. patient 220's upper-right measurements) and asymmetry in the individual distributions (e.g. densities on the right, without taking the temporal aspect into account).

\subsection{Model specification}\label{sec:app_model}

To fit these data, we considered the ALDRM defined in equation \eqref{eq:ALDRM} of section \ref{sec:aldrm} with:
\begin{equation}\label{eq:ALDRM_app}
\left\{
\begin{array}{rcl}
\mu_{ij}\; & = & \beta_0 + \beta_1 t_{ij} + \beta_2 t_{ij}^2 + \beta_3 Age_{i} + \beta_4 Sex_{i}  + b_{i0} + b_{i1} t_{ij} + b_{i2} t_{ij}^2 \\
log\left(\sigma_{ij}\right) & = & \xi_0 + \xi_1 t_{ij} + \xi_2 Age_{i}  + u_{i0} + u_{i1} t_{ij} \\
logit\left(\tau_{ij}\right) & = & \alpha_1 Age_{i} + \alpha_2 Sex_{i} + a_{i0} + a_{i1} t_{ij} \\
\end{array}
\right. \; ,
\end{equation}
where $\textbf{b}_i=\left(b_{i0},b_{i1},b_{i2} \right)^\top$, $\textbf{u}_i=\left(u_{i0},u_{i1} \right)^\top$, $\textbf{a}_i=\left(a_{i0},a_{i1} \right)^\top$ such as
\begin{equation*}
\textbf{r}_i =
\begin{pmatrix}
    \textbf{b}_i \\
    \textbf{u}_i \\
    \textbf{a}_i \\
\end{pmatrix}
\sim \mathcal{N} \left(
\begin{pmatrix} 
    \textbf{0}  \\
    \textbf{0}  \\
    \textbf{0}  \\
\end{pmatrix}
\; , \; 
\begin{pmatrix} 
    \Sigma_b & 0 & 0 \\
    0 & \Sigma_{u} & 0 \\
    0 & 0 & \Sigma_{a} \\
\end{pmatrix}
\right).
\end{equation*}
The covariance matrices $\Sigma_b$, $\Sigma_u$ and $\Sigma_a$ are assumed to be unstructured. Note that age variable is standardized to improve convergence, and the sex variable is equal to 1 for men and 0 for women. The model was selected by including only those covariates that showed an effect in the model components.

\subsection{Results}\label{sec:app_result}

Table \ref{tab:res_appli} presents the estimate of ALDRM defined in equation  \eqref{eq:ALDRM_app}. 
Blood pressure levels tended to increase and then decrease over the course of follow-up (location part). Systolic blood pressure level is higher for men than for women ($\beta_{Male}=0.632$ with a credible interval at $95\%$ of $CI_{95\%}=[0.17;1.09]$) and increases with age ($\beta_{Age}=0.61$ and $CI_{95\%}=[0.38;0.84]$).
The dispersion of measurements around the mode (scale part) tends to decrease with time ($\xi_{time}=-0.011$ and $CI_{95\%}=[-0.019; -0.003]$) and to increase with age ($\xi_{Age}=0.084$ and $CI_{95\%}=[0.054; 0.114]$). 
Finally, asymmetry around the mode seems to be affected by age and gender (skewness part). A man tends to have a more left-tailed distribution than a woman ($\alpha_{Male}=0.146$ and $CI_{95\%}=[0.041; 0.251]$) , i.e. a stronger dispersion for values below the mode than for values above. In other words, the estimate of the asymmetry parameter of the $\mathcal{AL}$ distribution is closer to 1 for a man than for a woman, all other things being equal. 
Finally, the older a person is, the more likely he/she is to have a distribution with a left heavy tail($\alpha_{Age}=0.126$ and $CI_{95\%}=[0.052; 0.199]$). 

\begin{table}[t!]
\caption{Estimation of ALDRM defined in equation \eqref{eq:ALDRM_app} including the posterior mean of parameters ($\widehat{\theta}$), their standard deviation of posterior sample ($sd$), the credible interval at $95\%$ ($CI_{95\%}(\theta)$) and the Gelman-Rubin convergence criteria ($\widehat{R}$). 
3 chains, each with 40,000 iterations, a burn-in period of 10,000 iterations, and a thinning interval of 10 are considered.}\label{tab:res_appli}
\begin{center}
\begin{tabular}{llcccc}
\hline 
&& $\widehat{\theta}$ & $sd$ & $CI_{95\%}(\theta)$ & $\widehat{R}$  \\ 
\hline
\multicolumn{6}{l}{\textbf{Location part}} \\  
& $\beta_{Intercept}$ & 13.240 & 0.163 & [12.922;13.559] & 1.000 \\
& $\beta_{time}$   & 0.310 & 0.041 & [0.231;0.392] & 1.000 \\
& $\beta_{time^2}$ & -0.030 & 0.004 & [-0.038; -0.021] & 1.000 \\
& $\beta_{Age}$ & 0.610 & 0.119 & [0.380; 0.842] & 1.000 \\
& $\beta_{Male}$   & 0.632 & 0.235 & [0.17; 1.091] & 1.000 \\
& $\Sigma_{b_{11}}$  & 3.049 & 0.334 & [2.456; 3.771] & 1.001 \\
& $\Sigma_{b_{22}}$  & 0.295 & 0.037 & [0.231; 0.375] & 1.001 \\
& $\Sigma_{b_{33}}$  & 0.003 & 0.0004 & [0.002; 0.004] & 1.000 \\
& $\Sigma_{b_{12}}$  & -0.324 & 0.084 & [-0.499; -0.173] & 1.002  \\
& $\Sigma_{b_{13}}$  &  0.014 & 0.009 & [-0.002; 0.032] & 1.001 \\
& $\Sigma_{b_{23}}$  & -0.025 & 0.004 & [-0.033; -0.018] & 1.000 \\
\hline
\multicolumn{5}{l}{\textbf{Scale part}} \\ 
& $\xi_{Intercept}$  & -0.555 & 0.019 & [-0.592; -0.517] & 1.022 \\ 
& $\xi_{time}$       & -0.011 & 0.004 & [-0.019; -0.003] & 1.008 \\
& $\xi_{Age}$        & 0.084 & 0.016  & [0.054;  0.114] & 1.002 \\
& $\Sigma_{u_{11}}$  & 0.059 & 0.008  & [0.044;  0.077] & 1.003 \\
& $\Sigma_{u_{22}}$  & 0.002 & 0.0003  & [0.001;  0.002] & 1.001 \\
& $\Sigma_{u_{12}}$  & -0.003 & 0.001 & [-0.006; -0.001] & 1.003 \\
\hline
\multicolumn{5}{l}{\textbf{Skewness part}} \\ 
& $\alpha_{Age}$    & 0.126 & 0.038 & [0.052;  0.199] & 1.003 \\
& $\alpha_{Male}$   & 0.146 & 0.053 & [0.041;  0.251] & 1.003 \\
& $\Sigma_{a_{11}}$ & 0.221 & 0.039 & [0.151;  0.304] & 1.002 \\
& $\Sigma_{a_{22}}$ & 0.012 & 0.002 & [0.010;  0.016] & 1.000 \\
& $\Sigma_{a_{12}}$ & -0.023 & 0.006 & [-0.036; -0.011] & 1.001 \\
\hline 
\end{tabular} 
\end{center}
\end{table}

\subsection{Goodness-of-fit}\label{sec:app_result}

To assess the goodness-of-fit to data, the ALDRM is compared to the LSLQMM (ALDRM assuming that $\tau_{ij}=0.5$) and the LSMM (assuming Gaussian distribution). 
To make the three models comparable, we considered a similar Bayesian estimation procedure, whether in terms of prior definition, or the MCMC (Markov Chain Monte Carlo) sampling settings (using 3 chains, each with 40,000 iterations, a burn-in period of 10,000 iterations, and a thinning interval of 10).

When considering the criterion $C_\Gamma$, none of the three models appears to be better than the others. Table \ref{tab:appl_criteria} presents the criteria calculated based on different sets of orders and loss functions (absolute or quadratic costs). 
Considering all deciles or the three quartiles, LSMM appears better than the two other models but considering the median and first and last decile ALDRM is better. This could suggest that ALDRM better fit extreme quantiles but not the central quantiles.
For data adjustment, no model seems better than the order, especially when comparing the values to those obtained in the simulations.
The distributions of the computed individual criteria $C_{i,\Gamma}$ are shown in Figure \ref{fig:fig_appli_box_criteria} according the models, the sets of quantiles and the loss function used.

\begin{table}[th]
    \caption{Estimated criterion $C_\Gamma$ defined in equation \eqref{eq:criterC} based on absolute cost loss function for MMAE and quadratic loss function for MMSE. Set 1 denoted by $\Gamma_1=\left\{0.1,0.2,\ldots,0.9\right\}$ including all decile orders, Set 2 denoted by $\Gamma_2=\left\{0.25,0.5,0.75\right\}$ including all quartile orders, and Set 3 denoted by $\Gamma_3=\left\{0.1,0.5,0.9\right\}$ including median, first and ninth deciles.}
    \label{tab:appl_criteria}
    \centering
    \begin{tabular}{lccccccc}
    \hline
       &        & MMAE   &        &&        & MMSE   &      \\
              \cline{2-4}                 \cline{6-8}
       & Set 1  & Set 2  & Set 3  && Set 1  & Set 2  & Set 3   \\
    \hline
ALDRM  & 0.0264 & 0.0311 & 0.0153 && 0.0012 & 0.0015 & 0.0005 \\
LSLQMM & 0.0275 & 0.0277 & 0.0167 && 0.0013 & 0.0015 & 0.0006 \\
LSMM   & 0.0226 & 0.0239 & 0.0191 && 0.0010 & 0.0011 & 0.0007 \\
    \hline
    \end{tabular}
\end{table}

The estimations of both LSLQMM and LSMM are shown in Table \ref{tab:app_aldrm_lsmm}. Concerning parameter estimates for the location component, the three models are very close. This suggests that the data distribution is symmetrical at the population level, and that the mean, median and mode are therefore equivalent. Comparing estimates of the scale component is more difficult, as this component does not define the same quantity of the distribution, and it is directly link to the skewness component. This is reflected in the intercept estimate, which differs between the two models. But we find the same trends in covariates (whether time or age) in the dispersion of the data. Finally, ALDRM has the advantage of also explaining the asymmetry of the distribution based on covariates. \\

\begin{figure}[!t]
\begin{center}
\includegraphics[scale=0.9]{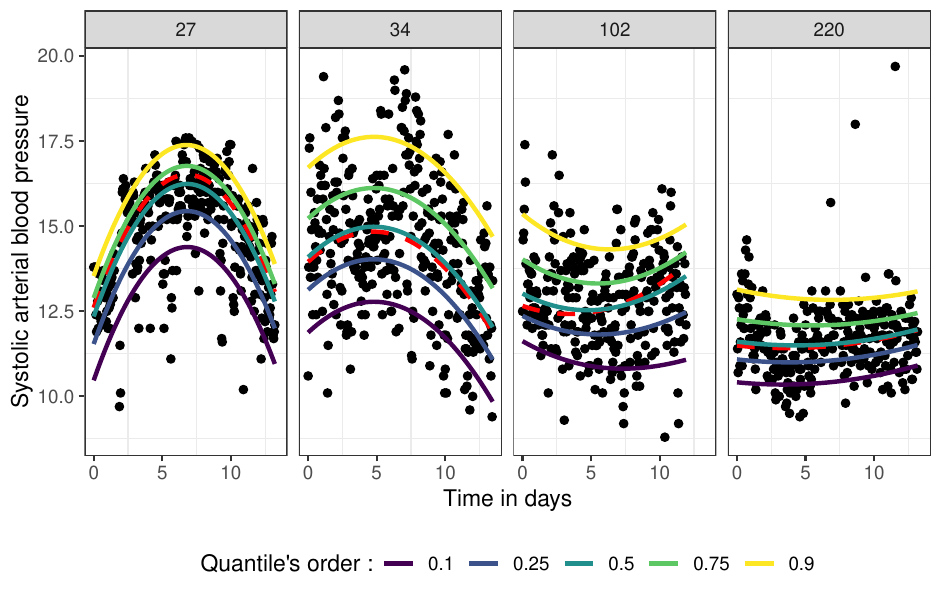}
\end{center}
\caption{Longitudinal measurements (black dots) and predicted conditional quantile trajectories (solid lines) using equation \eqref{eq:ALD_quantile_fct}, and mode trajectory (dotted line in red, corresponding at $\widehat{\mu}_i(t)$ given predicted random effects $\widetilde{\boldsymbol{b}_i}$) for four selected individuals.}\label{fig:id_points_mode_quantile}
\end{figure}

Regarding the individual fit of the ALDRM to the data, Figure \ref{fig:id_points_mode_quantile} represents the evolution of the individual distributions for four subjects through its mode (in dotted red) and five quantiles (the three quartiles, and the first and ninth deciles). 
For patient 27, we note that the blood pressure increase over the first 7 days and then decrease over the last 7 days. The blood pressure interval between the third quartile and the median is smaller than that between the median and the first quartile at all follow-up times. This suggests that the measurements for this individual are more dispersed for low values than for high blood pressure values (heavy tail to the left). This information is reinforced by the trend of the mode, which is always higher than that of the median, leading that the asymmetry parameter for this individual is always greater than $0.5$. Finally, we expect to observe $10\%$ of observations below the trajectory of the first decile (or above the ninth), and for a fixed time, our model predicts the probability of observing a measurement below the first decile (or above the ninth) to be $0.1$.

For patient 34, the same trend can be seen in measurements over time, with an increase on the first days and then a decrease. However, there is greater variability between measurements, characterized by more widely spaced quantile trajectories. The distribution of measurements for this individual appears relatively symmetrical, since the modal trajectory is almost identical to the median trajectory (asymmetry parameter constant over time, and almost equal to 0.5).

For patient 102, the trend is reversed compared to the first two, with a decrease over the first days and an increase over the following days. Interetingly, the model is able to capture the asymmetry of the data as it changes over the course of follow-up. At the start of the follow-up, the measurements showed a strong dispersion for high blood pressure measurements (heavy tail to the right, with an asymmetry parameter below 0.5), whereas a strong dispersion for low measurements is observed and predicted at the end of the follow-up (heavy tail to the left, with an asymmetry parameter above 0.5).

For patient 220, the measurements are relatively constant over time and lower than those of the first three, with less dispersion. The dispersion decreases over time, although some extreme values are observed at the end of the follow-up. This illustrates the robustness of the model to extreme values. 

Note that Figure \ref{fig:id_points_mode_quantile} help to understand the criterion $C_{i,\Gamma}$ given in equation \eqref{eq:criterCi}. For a specific quantile order $\gamma\in]0,1[$, $\widehat{\gamma}_i$ corresponds to the proportion of measurements below the individual trajectory associated with the quantile order $\gamma$ for the $i$th individual. \\

A more complete representation of the evolution of the individual conditional distribution is given in Figure \ref{fig:density_cum_func}. Using the predictions of individual mode $\widetilde{\mu}_i(t)$, individual scale $\widetilde{\sigma}_i(t)$ and individual skewness $\widetilde{\tau}_i(t)$ given the predicted random effects $\widetilde{\boldsymbol{r}}_i$, it is possible to explicitly compute the probability density $\widehat{f}_{Y_i(t)|\widetilde{\boldsymbol{r}}_i}(v)$ at any time $t$ and for any blood pressure value $v$. Panel A of Figure \ref{fig:density_cum_func} fully illustrates the dispersion around the mode of the individual distribution, without restricting ourselves to a few quantiles. It is a kind of density map defined as a function of time for given arterial pressure values.

Similarly, the cumulative distribution function $F_{Y_i(t)|\textbf{r}_i}(v)=Pr\left(Y_i(t)\leq v \vert \textbf{r}_i \right)$ is directly estimable given parameter estimations and predictions of random effects from the formula given in equation \eqref{eq:ALD_cum_distr}. Panel B of Figure \ref{fig:density_cum_func} fully illustrates the probability of observing blood pressure values below a certain threshold (blood pressure value), and for a given time. 

\section{Discussion}

To fit intensive heterogeneous longitudinal data, we have proposed a new mixed-effects distributional regression model based on $\mathcal{AL}$ distribution.
Standing at the intersection of quantile regression and distributional regression, the model we propose offers several key advantages. 
It enables a flexible fitting of individual distributions, while accounting for inter-individual heterogeneity in marker dynamics in terms of heteroskedasticity and asymmetry of measurements over time. Furthermore, this generalization of the linear quantile mixed-effects model inherits its robustness to extreme values, avoiding estimation biases that could affect the evolution of the mean. 
Although our parametric approach is less flexible than a non-parametric approach, it has the advantage of being easily interpretable, especially since the parameters of the $\mathcal{AL}$ distribution have an intuitive and direct interpretation of the shape of the distribution. Indeed, the location parameter is the mode of the distribution, the scale parameter quantifies the dispersion around the mode, and the skewness parameter indicates the asymmetry around the mode. In the proposed ALDRM,  three parameters are modeled using a linear predictor that includes covariates and random effects. In this way, it is easy to study the impact of covariates on each of these components of the biomarker distribution, and their evolution over time, whether at population or individual level, thanks to the introduction of subject-specific random effects.
The simulation study validated the Bayesian estimation procedure that we implemented in the \texttt{R} package \texttt{BeQut}. 
Furthermore, although the proposed model includes numerous random effects and parameters, the quality of the model's estimation is not affected by the number of repeated measurements per subject, whether there are few measurements (only 10 measurements) or intensive measurements (200 repeated measurements). Similarly, the number of subjects does not affect the quality of the model estimation.

In order to choose between two distributional regression models that may differ in terms of the assumed distribution of the data, or in terms of the assumption made about the shape of the individual distributions, we have also proposed a new model selection criterion for longitudinal data. It measures the proximity between the individual distribution estimated by the model and the empirical individual distribution of the data over time, using a set of quantiles. The simulation study showed that this criterion works well regardless of the number of repeated measurements, but it discriminates better as the number of repeated measurements increases. \\

Modeling the distribution of blood pressure measurements of patients hospitalized in intensive care units illustrated and highlighted the advantages of the proposed model. 
Although the $\mathcal{AL}$ distribution is not widely used in practice to fit data, the application results demonstrate its relevance for data modeling.
Indeed, the goodness-of-fit of the distributional regression mixed model based on the $\mathcal{AL}$ distribution is comparable to that based on the Gaussian distribution. 
The ALDRM allows for a much more detailed description and understanding of the data by modeling the location, dispersion, and asymmetry of the distribution through explanatory variables. The inclusion of subject-specific random effects on each of these components of the distribution also allows for the adjustment of the individual distribution over time. 
Finally, ALDRM offers other advantages for data analysis. Indeed, the probability of observing a measurement within a specific interval has an explicit form depending on the parameters, because the cumulative distribution function of the $\mathcal{AL}$ distribution is explicit. Furthermore, any quantile of the distribution also has an explicit form given its parameters. Consequently, the study and prediction of multiple quantiles of the distribution over time does not lead to crossing issues, unlike a classical quantile regression approach.

\section*{Acknowledgments}

Computer time for this article was provided by the computing facilities MCIA of the \textit{Université de Bordeaux} and of the \textit{Université de Pau et des Pays de l'Adour}.

\section*{Funding}

A part of this work was funded by the French National Research Agency (grant ANR-21-CE36 for the project "Joint Models for Epidemiology and Clinical research").

\bibliographystyle{apalike}
\bibliography{biblioTex}

\newpage

\appendix

\renewcommand\thefigure{\thesection\arabic{figure}}
\renewcommand\thetable{\thesection\arabic{table}}

\section{Location-scale linear mixed model}\label{sec:lsmm}

LSMM is a basic mixed-effects distributional regression model based on the Gaussian distribution assumption. 
In this section, we use the same notation as in the main text, to keep things as consistent as possible.
Let $\boldsymbol{Y_i}=(Y_{i1},\ldots,Y_{in_i})^{\top}$ denotes the vector of the $n_i$ measurements for the group $i(i=1,\ldots,n)$. The response variable $Y_{ij}$ is assumed to be distributed from the Gaussian distribution conditionally to group-specific random effects $\textbf{r}_i$ such as
\begin{equation}\label{eq:lsmm}
Y_{ij}\vert \textbf{r}_i \sim \mathcal{N}\left( \mu_{ij}, \sigma_{ij} \right)
\quad \text{with} \quad
\left\lbrace 
\begin{array}{lcl}
\mu_{ij} & = & \textbf{x}_{ij,\mu}^\top\boldsymbol{\beta} \; + \;  \textbf{z}_{ij,\mu}^\top \textbf{b}_{i} \\
\log\left(\sigma_{ij}\right) & = & \textbf{x}_{ij,\sigma}^\top\boldsymbol{\xi} \; + \;  \textbf{z}_{ij,\sigma}^\top \textbf{u}_{i} 
\end{array}
\right.
\end{equation}
where the location $\mu_{ij,\tau}$ and the scale $\sigma_{ij,\tau}$ can be defined as a function of a linear predictor composed of both fixed and random effects. 
Classically, we assume independence between the location part and the scale part (residual space), leading us to assume the following distribution of individual random effects: 
\begin{equation}
\textbf{r}_i =
\begin{pmatrix}
    \textbf{b}_i \\
    \textbf{u}_i \\
\end{pmatrix}
\sim \mathcal{N} \left(
\begin{pmatrix} 
    \textbf{0}  \\
    \textbf{0}  \\
\end{pmatrix}
\; , \; 
 \Sigma_r = \begin{pmatrix} 
    \Sigma_b & 0  \\
    0 & \Sigma_{u}  \\
\end{pmatrix}
\right).
\end{equation}

Although this model is already implemented in the \texttt{lsmm} function of the \texttt{LSJM} R-package available on GitHub (\url{https://github.com/LeonieCourcoul/LSJM}), the frequentist framework used has two drawbacks in our work: the computation time generated by the intensive measurements to model, and the comparison with ALDRM which is implemented in a Bayesian framework.
We therefore consider a Bayesian estimation procedure like the one considered in section \ref{sec:estimation}. The posterior distribution of parameters is:
\begin{equation*}\label{eq:posterior_distrib}
\pi\left(\boldsymbol{\theta},\boldsymbol{r}\vert \boldsymbol{y}\right) \propto \prod^{n}_{i=1} \prod^{n_i}_{j=1} f\left( y_{ij} \vert \textbf{r}_{i}, \boldsymbol{\theta} \right) \phi\left( \textbf{r}_{i}\vert\Sigma_r\right) \pi\left(\boldsymbol{\theta}\right) 
\end{equation*}
where $f\left(.\right)$ is a Gaussian density function associated with distributions defined in \eqref{eq:lsmm}, $\phi$ denotes the zero-mean normal density function for subject-specific random effects and $\pi\left(\boldsymbol{\theta}\right)$ is the prior distribution of parameters that are assumed to be independent.  
For prior distributions of parameters, normal distributions for regression parameters and inverse Wishart ($\mathcal{IW}$) distributions for the covariance matrices of random effects are considered: 
\begin{equation*}
\boldsymbol{\beta} \sim \mathcal{N}_{p_\beta} \Big( \boldsymbol{\mu}_\beta , \; \Sigma_\beta \Big), 
\quad \boldsymbol{\xi} \sim \mathcal{N}_{p_\sigma} \Big( \boldsymbol{\mu}_\xi , \; \Sigma_\xi \Big), 
\quad \Sigma_b \sim \mathcal{IW}\Big(\boldsymbol{\omega}_b, \; \Omega_b \Big) \text{ and }
\Sigma_u \sim \mathcal{IW}\Big(\boldsymbol{\omega}_u, \; \Omega_u \Big),
\end{equation*}
These priors are considered as vague to estimate the LSMM as well as the ALDRM. 
The estimation of models is implemented in the function \texttt{lsmm} of the \texttt{R}-package \texttt{BeQut} and statistical inference is based on posterior parameter samples. Bayesian Markov chain Monte Carlo (MCMC) sampling is performed using the \texttt{JAGS} software \citep{plummer_2016}. The open source \texttt{R}-package \texttt{BeQut} is available on Github at the following link: \url{https://github.com/AntoineBbi/BeQut}.

\newpage 

\section{Simulation complements}

\begin{figure}[!h]
\begin{center}
\includegraphics[scale=0.8]{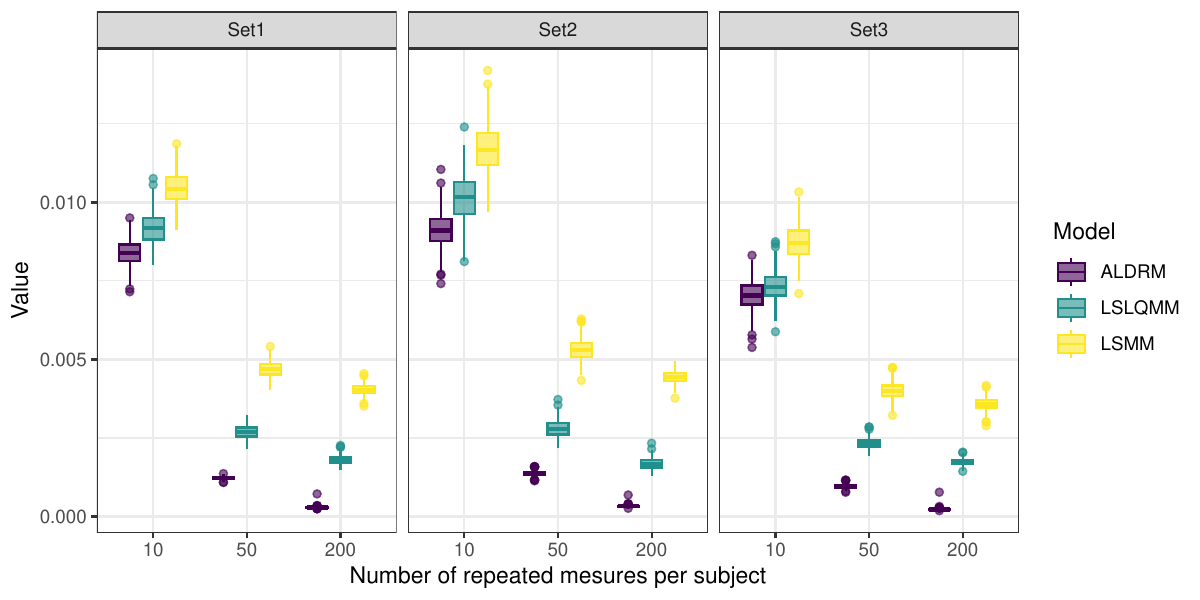}
\end{center}
\caption{Distribution of the $N=500$ computed criteria $C_\Gamma$ proposed in equation \eqref{eq:criterC} based on quadratic cost loss function $\ell\left(x\right)=x^2$ for all $x\in\mathbb{R}$, the MMSE. Scenarii with $n=200$ and $m=10,50,200$ are considered. Set 1 denoted by $\Gamma_1=\left\{0.1,0.2,\ldots,0.9\right\}$ including all decile orders, Set 2 denoted by $\Gamma_2=\left\{0.25,0.5,0.75\right\}$ including all quartile orders, and Set 3 denoted by $\Gamma_3=\left\{0.1,0.5,0.9\right\}$ including median, first and ninth deciles.}\label{fig:boxplot_criteria_MMSE}
\end{figure}

\begin{figure}[!h]
\begin{center}
\includegraphics[scale=0.8]{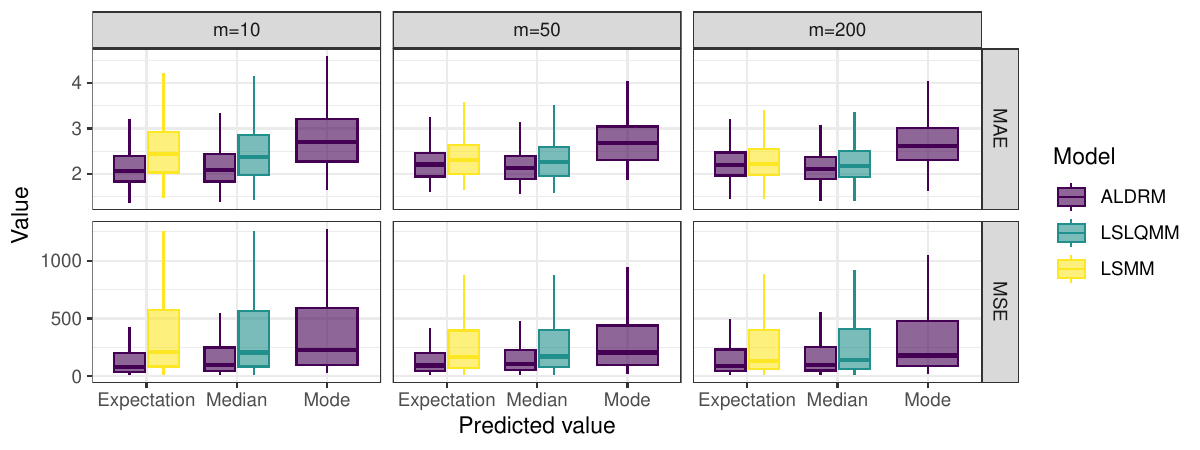}
\end{center}
\caption{Distribution of the MSEs and MAEs for simulation scenario with $N=500$, $n=200$ and $m=10,50,200$. $MSE=\frac{1}{n m} \sum_{i=1}^n\sum_{j=1}^{m} (y_{ij}-\widehat{y}_{ij})^2$ and $MAE=\frac{1}{nm} \sum_{i=1}^n\sum_{j=1}^{m} \vert y_{ij}-\widehat{y}_{ij}\vert$ where $\widehat{y}_{ij}$ is the predicted value. For the LSMM, $\widehat{y}_{ij}=\widehat{\mu}_{ij}$ is the predicted expectation of the individual distribution given random effects. For the LSLQMM, $\widehat{y}_{ij}=\widehat{\mu}_{ij}$ is the predicted median given random effects. For the ALDRM, $\widehat{y}_{ij}=\widehat{\mu}_{ij}$ is the predicted mode of the distribution, and the predicted expectation $\widehat{y}_{ij}=\widehat{\mathds{E}}(Y_{ij}\vert \tilde{\textbf{r}}_i,\widehat{\theta})$ and the predicted median $\widehat{y}_{ij}=\widehat{\mathcal{Q}}_{Y_{ij}\vert \tilde{\textbf{r}}_i,\widehat{\theta}}(0.5)$ are also considered. }\label{fig:boxplot_errors_m}
\end{figure}

\newpage

\section{Application complements}\label{sec:res_aldrm_lsmm}

\begin{figure}[!h]
\begin{center} 
\includegraphics[scale=0.85]{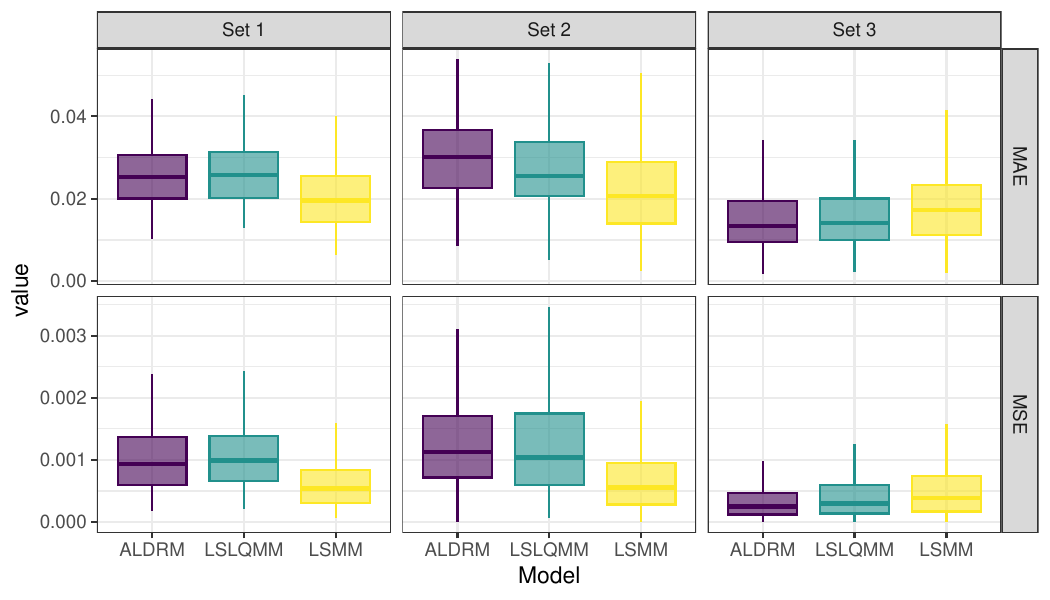}
\end{center}
\caption{Distribution of criterion $C_{i,\Gamma}$ defined in equation \eqref{eq:criterC} based on absolute loss function for MMAE and quadratic loss function for MMSE. Set 1 denoted by $\Gamma_1=\left\{0.1,0.2,\ldots,0.9\right\}$ including all decile orders, Set 2 denoted by $\Gamma_2=\left\{0.25,0.5,0.75\right\}$ including all quartile orders, and Set 3 denoted by $\Gamma_3=\left\{0.1,0.5,0.9\right\}$ including median, first and ninth deciles.}\label{fig:fig_appli_box_criteria}
\end{figure}

\begin{table}[!h]
\caption{Estimation of the LSLQMM for the median ($\tau=0.5$) and the LSMM for the expectation including the posterior mean  of parameters ($\widehat{\theta}$), their standard deviation of posterior sample ($sd$), the credible interval at $95\%$ and the Gelman-Rubin convergence criteria ($\widehat{R}$). }\label{tab:app_aldrm_lsmm}
\begin{center}
\scalebox{0.9}{
\begin{tabular}{llcccc|cccc}
\hline \\
\multicolumn{2}{l}{} & \multicolumn{4}{c|}{\texttt{LSLQMM ($\tau=0.5$)}}  & \multicolumn{4}{c}{\texttt{LSMM}}   \\
 && $\widehat{\theta}$ & $sd$ & $CI_{95\%}(\theta)$ & $\widehat{R}$  &  $\widehat{\theta}$ & $sd$ & $CI_{95\%}(\theta)$ & $\widehat{R}$ \\ 
\hline
\multicolumn{6}{l}{\textbf{Location part}} \\  
& $\beta_{Intercept}$& 
13.227 & 0.147 & [12.933; 13.515]& 1.000 & 
13.236 & 0.140 & [12.962;13.515] & 1.000 \\
& $\beta_{time}$     & 
0.309 & 0.041 & [0.229;0.392] & 1.000 & 
0.307 & 0.041 & [0.226;0.389] & 1.000 \\
& $\beta_{time^2}$   & 
-0.030 & 0.005 & [-0.039;-0.021] & 1.000 & 
-0.030 & 0.005 & [-0.039;-0.021] & 1.000 \\
& $\beta_{Age}$      & 
0.470 & 0.106 & [0.262;0.681] & 1.000 & 
0.425 & 0.101 & [0.229;0.625] & 1.000 \\
& $\beta_{Male}$     & 
0.500 & 0.208 & [0.098;0.913] & 1.000 & 
0.463 & 0.202 & [0.073;0.860] & 1.000 \\
& $\Sigma_{b_{11}}$  & 
2.481 & 0.263 & [2.019;3.044] & 1.000 & 
2.212 & 0.233 & [1.799;2.715] & 1.000\\
& $\Sigma_{b_{22}}$  & 
0.300 & 0.036 & [0.237;0.376] & 1.000 & 
0.286 & 0.035 & [0.224;0.363] & 1.000\\
& $\Sigma_{b_{33}}$  & 
0.003 & 0.000 & [0.003;0.004] & 1.000 & 
0.003 & 0.000 & [0.003;0.004] & 1.000\\
& $\Sigma_{b_{12}}$  & 
-0.310 & 0.074 & [-0.464;-0.176] & 1.000 & 
-0.254 & 0.069 & [-0.403;-0.130] & 1.000\\
& $\Sigma_{b_{13}}$  &  
0.019 & 0.008 & [0.004;0.035] & 1.000 & 
0.016 & 0.008 & [0.002;0.031] & 1.000\\
& $\Sigma_{b_{23}}$  & 
-0.026 & 0.004 & [-0.034;-0.020] & 1.000 & 
-0.026 & 0.004 & [-0.034;-0.019] & 1.000\\
\hline
\multicolumn{5}{l}{\textbf{Scale part}} \\ 
& $\xi_{Intercept}$  & 
-0.506 & 0.023 & [-0.553;-0.458] & 1.016 & 
 0.437 & 0.022 & [ 0.398; 0.480] & 1.091 \\ 
& $\xi_{time}$       & 
-0.007 & 0.004 & [-0.015; 0.001] & 1.009 & 
-0.009 & 0.004 & [-0.017;-0.001] & 1.025 \\
& $\xi_{Age}$        &  
0.081 & 0.017 & [0.046; 0.116] & 1.102 & 
0.076 & 0.017 & [ 0.046; 0.112] & 1.054 \\
& $\Sigma_{u_{11}}$  &  
0.066 & 0.008 & [0.051;0.085] & 1.002 & 
0.075 & 0.017 & [0.046;0.112] & 1.001\\
& $\Sigma_{u_{22}}$  &  
0.002 & 0.000 & [0.002;0.003] & 1.001 & 
0.002 & 0.000 & [0.002;0.003] & 1.000\\
& $\Sigma_{u_{12}}$  & 
-0.004 & 0.001 & [-0.006;-0.002] & 1.004 & 
-0.006 & 0.001 & [-0.008;-0.003] & 1.000 \\
\hline
\end{tabular} 
}
\end{center}
\end{table}

\begin{figure}[!ht]
\begin{center}
\includegraphics[scale=0.9]{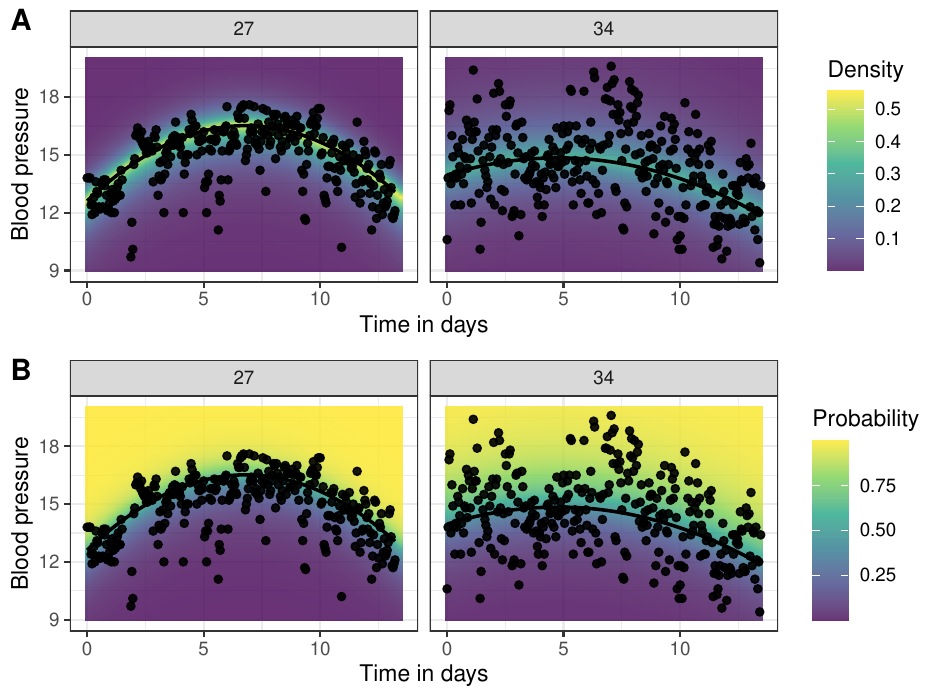} 
\end{center}
\caption{Maps of individuals 27 and 34 for estimated values from ALDRM of conditional density $\widehat{f}_{Y_i(t)|\widetilde{\boldsymbol{r}}_i}(v)$ (panels A) and conditional cumulative probability $\widehat{F}_{Y_i(t)|\widetilde{\boldsymbol{r}}_i}(v)$ (panel B) given blood pressure values (v) and time (t). The black trajectory represents the mode of the individual distribution over time.}\label{fig:density_cum_func}
\end{figure}

\end{document}